\documentclass[11pt]{iopart}
\usepackage{iopams}
\expandafter\let\csname equation*\endcsname\relax
\expandafter\let\csname endequation*\endcsname\relax
\usepackage{amsmath,amssymb,amsfonts,amsthm}
\usepackage{psfrag,latexsym}
\usepackage{amssymb}
\usepackage{mathtools}
\usepackage{physics}
\usepackage{cancel} 
\usepackage{bm}
\usepackage{dcolumn}
\usepackage{epic} 
\usepackage{epsfig}
\usepackage{feynmp}
\usepackage{graphicx}
\usepackage{grffile}
\usepackage[usenames, dvipsnames]{xcolor} 
\usepackage[breaklinks,colorlinks = true,linkcolor = red,urlcolor=blue,citecolor=cyan]{hyperref}
\usepackage{caption}
\usepackage{subcaption}
\usepackage{soul}
\usepackage{wrapfig}
\usepackage{xy} 
\usepackage{xcolor}
\usepackage{tikz}

\usepackage{makecell}

\usepackage{cite}
\usepackage{textcomp}
\usepackage{bbold}
\usepackage[scr=boondox,  
            cal=boondoxo]   
           {mathalpha}
\usepackage{ulem}
\usepackage[mathscr]{eucal}
\usepackage[ascii]{inputenc}

\usepackage{etoolbox}
\makeatletter
\newrobustcmd{\fixappendix}{%
  \patchcmd{\l@section}{1.5em}{7em}{}{}%
  \patchcmd{\l@subsection}{2.3em}{7em}{}{}%
}
\makeatother

\usepackage{color}
\usepackage{dsfont}

\definecolor{dgreen}{rgb}{0,0.7,0}

\newcommand{\mc}[1]{\mathcal{#1}}  
\newcommand{\msc}[1]{\mathscr{#1}}  
  
\newcommand{\mf}[1]{\mathfrak{#1}} 
\newcommand{\mt}[1]{\mathtt{#1}} 
\newcommand{\mfb}[1]{\boldsymbol{\mathfrak{#1}}}

\def\mtX{\mathtt{X}}
\def\mtt{\mathtt{t}}
\begin{document}

\title{Microscopic and hydrodynamic correlation in $1$d hard rod gas}
\author{Indranil Mukherjee}
\address{International Centre for Theoretical Sciences, TIFR, Bengaluru -- 560089, India}
\ead{indranil.mukherjee@icts.res.in}
\author{Seema Chahal}
\address{International Centre for Theoretical Sciences, TIFR, Bengaluru -- 560089, India}
\ead{seema.s@icts.res.in}
\author{Anupam Kundu}
\address{International Centre for Theoretical Sciences, TIFR, Bengaluru -- 560089, India}
\ead{anupam.kundu@icts.res.in}
\begin{abstract}
We compute  mass density correlations of a one-dimensional gas of hard rods at both microscopic and macroscopic scales. 
We provide exact analytical calculations of the microscopic correlation. For the correlation at macroscopic scale, we utilize Ballistic Macroscopic Fluctuation Theory (BMFT) to derive an explicit expression for the correlations of a coarse-grained mass density, which reveals the emergence of long-range correlations on the Euler space-time scale. By performing a systematic coarse-graining of our exact microscopic results, we establish a micro-macro correspondence and demonstrate that the resulting macroscopic correlations agree precisely with the predictions of BMFT. This analytical verification provides a concrete validation of the underlying assumptions of hydrodynamic theory in the context of hard rod gas.
\end{abstract}
\maketitle

\tableofcontents
\markboth{\it Microscopic and hydrodynamic correlation in $1$d hard rod gas}{}

\section{Introduction}
A fundamental objective in the statistical mechanics of many-body systems is to characterize the structure of correlations for the observables of physical interest. Such correlations can be either static or involve space-time dependence, including fluctuations of local observables. In equilibrium one-dimensional interacting particle systems, the analysis of space-time correlations generated by Hamiltonian dynamics provides key insights into transport properties. For non-integrable systems, which possess only a small number of conserved quantities, space-time correlation functions in one dimension exhibit universal features that are successfully described within the framework of fluctuating hydrodynamics of the conserved densities \cite{VanBeijeren2012,spohn2014nonlinear, Narayan2002}. In contrast, integrable systems, characterized by an extensive set of conservation laws, lead to non-universal space-time correlations, and only limited analytical results are currently available, largely relying on hydrodynamic approaches. For the integrable system, the Euler-scale GHD, together with its leading corrections, has been proven highly effective in describing the evolution of the conserved densities, where the GHD equations are formulated in terms of the evolution of the quasiparticle densities \cite{doyonlecturenotes}. However, this Euler GHD framework is insufficient to account for the fluctuations and correlations in the system. It requires going beyond the description of average conserved densities and developing a statistical theory of macroscopic fluctuations. 

In conventional hydrodynamics, this can be achieved through Landau-Lifshitz theory \cite{landau1987fluid} in which one adds phenomenological stochastic forces to each dissipative current in the linear response regime -- leading to linear fluctuating hydrodynamic equations \cite{de2006hydrodynamic, de2013non}. However, such a linearised HD theory fails to describe transport phenomena in certain cases, such as anomalous transport in low-dimensional systems, for which one requires an extension to non-linear fluctuating HD theory \cite{spohn2014nonlinear,spohn2016fluctuating}. For diffusive systems, macroscopic fluctuation theory, rooted in large deviation principles, provides a systematic characterization of the probabilities of density and current fluctuations on diffusive space-time scales \cite{bertini2015macroscopic, bertini2002macroscopic}.  Recent progress has extended these ideas to integrable systems, leading to a deeper understanding of correlation functions associated with conserved densities. In particular, for integrable models displaying ballistic transport, a large-deviation-based framework known as ballistic macroscopic fluctuation theory (BMFT) has been developed \cite{doyon2023emergence, doyon2023ballistic, AnupamMFT}. This approach successfully captures density and current fluctuations in the ballistic regime and predicts the emergence of long-range correlations even in systems with short-range interactions.

In general, it is difficult to derive hydrodynamic theory of an interacting many-particle system starting from a microscopic description. Often it is based on the assumption of hydrodynamic projection in which one projects the local microscopic-scale currents associated to the conserved charges on the space of conserved charges and expresses them as functions/functionals of these densities through the assumption of local equilibrium. In BMFT, this further requires the assumption of local relaxation. This assumes that fluctuations of coarse-grained observables on the Euler scale occur solely as functions of local conserved densities - as fixed by equilibrium averages - rather than independently \cite{doyon2023ballistic}. While these assumptions yield elegant descriptions of large-scale motion and fluctuations, their detailed predictions must be tested against microscopic analytical or numerical computations, which remain scarce in the literature. Recently, a few numerical tests of hydrodynamic solutions have been conducted for non-integrable systems using low-entropy initial conditions \cite{chakraborti2021blast, ganapa2021blast,singh2023blast, kumar2025shock, kumar2025splash,mukherjee2025extreme}, and similar verifications of Euler-scale density profiles and correlations have been performed for integrable systems \cite{doyon2023ballistic,doyon2023emergence}.

In this paper, we test the predictions of Generalized Hydrodynamics (GHD) and BMFT for mean density profiles and correlations, respectively, within a one-dimensional gas of hard rods. This integrable system consists of extended objects of length $a$ and unit mass interacting via hard-core repulsion. We analytically compute the space-time correlation of the mass density both microscopically and hydrodynamically using BMFT. While BMFT predicts correlations of coarse-grained observables describing hydro-scale fluctuations, the microscopic approach provides correlations at the micro-scale. However, these two results should be related through coarse-graining. By computing the correlations via both methods, we verify this  relation and thus test the validity of the hydrodynamic assumptions.

The paper is organised as follows. (i) In Sec.~\ref{sec:2}, we introduce the microscopic dynamics of hard rods and discuss their mapping to the hard-point gas model. The initial conditions considered in this work are described in this section. In Sec.~\ref{sec:den},  we  revisit the computation of the mass density for hard rods using the microscopic approach and discuss its extension in the thermodynamic limit, as derived in \cite{Mrinal_2024_HR}. (ii) In Sec.~\ref{sec:3}, we present the microscopic approach for computing the unequal space-time and equal-time correlations of the mass density. For the equal-time correlation, we demonstrate how the Euler-scale solution emerges from the microscopic description in the thermodynamic limit, and further discuss the extension of the microscopic solution for a large but finite number of rods, which provides an expression of the microscopic dynamics that goes beyond the Euler solution. (iii)  In Sec~\ref{sec:4}, we compute the space-time correlation of mass density in hydro-scale using BMFT and show how a systematic coarse-graining of the microscopic solution enables a direct comparison between microscopic and macroscopic observables. We demonstrate quantitative agreement between the BMFT prediction and the coarse-grained microscopic results. (v) Finally, Sec.~\ref{sec:5} summarizes our findings and outlines possible the future directions of this work. Supplementary derivations are provided in the appendix.

\section{Hard rods system, initial conditions and summary of results}\label{sec:2}

We consider a system of $N$ identical hard rods, each having unit mass and length $a$, moving ballistically on a straight line of infinite length. The positions and velocities of the hard rods are represented by $\{X_i, v_i\}$ for $i=1,2,\dots,N,$ where $X_i$ represents the ordered positions $(X_i<X_{i+1}-a)$ and $v_i$ the velocities.  These rods undergo ballistic motion between instantaneous elastic collisions, and during each collision, their momenta get exchanged. Clearly, in the limit $a=0,$ {\it i.e.} when the rod length approaches zero, the interacting hard-rod system reduces to a gas of hard-point particles (HPG), which through Jepsen mapping \cite{jepsen1965dynamics} can be further transformed to a system of non-interacting point particles.  The microscopic dynamics of hard rods can be effectively mapped onto a system of hard point particles through a specific transformation as follows \cite{percus1969exact,bernstein1988expansion,lebowitz1968time}. Starting with the configuration of rods $\{X_i,v_i\},$ one can construct an equivalent representation of  point particles $\{x_i,v_i\}$ 
using the following mapping 
\begin{equation}\label{eq:HR to HP mapping infinite line}
    x_i =X_i - (i-1)a, ~~\text{for} ~ i=1,2,\dots, N.
\end{equation}
In this point-particle framework, particles undergo purely ballistic motion. Unlike hard rods, they exchange their velocities during collisions without experiencing any jump in their positions, allowing them to be regarded as effectively non-interacting particles \cite{jepsen1965dynamics}. 

In this paper, we consider a particular choice of initial condition. Because every hard-rod configuration $\{X_i,v_i\}$ corresponds  to a unique configuration $\{x_i,v_i\}$ of  point particles, we first specify the initial state in the point-particle framework. The positions of $N$ point particles $\{x_i'\}$ are are drawn independently and identically from a  distribution $\phi(x')$ within the interval $[L_1,L_2]$, normalised such that $\int_{L_1}^{L_2} dx~\phi(x)=1$, and then ordered as $\{x_i\}=$ Order[$\{x_i'\}$] with $L_1<x_1<x_2<\dots<x_N<L_2$. Each particle $\{x_i\}$ is subsequently assigned a velocity $v_i,$ sampled independently from the normalized distribution $h(v_i).$ The joint distribution of positions and velocities for $N$ hard point particles can thus be expressed as
\begin{align}
 \mathbb{P}(\{x_{i},v_{i}\},0)=N!\prod_{i=1}^{N} \phi(x_{i})h(v_{i})~\prod_{i=1}^{N-1}\Theta(x_{i+1}-x_i)
 ,\label{eq:prob_init}
 \end{align}
where $\Theta(x)$ is the Heaviside theta function, and the product over Heaviside $\Theta$ functions ensures the ordering $\{x_{i+1}\ge x_{i}~;~i=1,...,N\}$. The initial mean phase-space density (PSD) and mean mass density of hard-point particles are given by $\bar {\mc f}(x,v)=N\phi\left(x\right)h(v)$ and $\bar \rho(x)=N\phi\left(x\right)$, respectively. We then take the thermodynamic limit $N\to \infty,$ and box size $(L_2-L_1)\to \infty$ ensuring that the initial mass density profile $\bar \rho(x)$ remains finite at all positions. In particular we choose $\phi(x)=\phi_0\left(\frac{x}{\sigma}\right)$ such that $\bar \rho(x)$ varies over a length scale $\sigma$. For each configuration of the point particles $\{x_i,v_i\}$, we subsequently map it to the hard rod coordinate $\{X_i,v_i\}$ following the inverse mapping of Eq. \eqref{eq:HR to HP mapping infinite line} {\it i.e.} $X_i=x_i+a(i-1)$.  This initial ensemble is particularly suitable for both microscopic analytical calculations and hydrodynamic calculations. In the literature, another type of initial ensemble is considered where the positions and velocities of the hard-rods are chosen from factorized distribution respecting the non-overlapping condition $X_{i+1}>X_i+a,~i=1,2,...,N$ \cite{doyon2023ballistic, AnupamMFT}. In this case the hard-rod gas initially has only short-range correlation unlike our case where the gas already contains long-range correlation (produced by the transformation in Eq.~\eqref{eq:HR to HP mapping infinite line}) to start with.

For the hard rod gas, the mean mass density profile ($\bar \varrho(X)$) is expressed in terms of the point-particle mass density ($\bar \rho(x)$):
\begin{align}
    \bar \varrho\big( X(x)\big)=\frac{\bar \rho(x)}{1+a\bar \rho(x)},
    \label{bar-varrho-ini}
\end{align}
with 
$X(x)=x+a\int dy~\Theta(x-y)\bar \rho(y)$.  This transformation is essentially the same mapping as described in Eq.~\eqref{eq:HR to HP mapping infinite line}, now expressed using the point-particle mass density $\bar \rho(x).$ It is easy to realize that for large $N$ and small $a$ with $\frac{Na}{\sigma}$ fixed, the average mass density $\bar \varrho(X)$ varies over a length scale $\ell \sim \sigma$. For all numerical simulations in this paper, $h(v)$ is chosen to be a Maxwell distribution $h(v)=\frac{1}{\sqrt{2\pi T}}\exp\left(-\frac{v^2}{2T}\right)$ with zero mean and temperature $T.$

In this paper, we analytically compute the  space-time correlation of mass density of hard rod gas both at microscopic  and macroscopic scales. The correlation at micro-scale is defined in terms of the  empirical mass density 
\begin{align}
   \hat{\varrho}(X,t)=\sum_{i=1}^N \delta\left (X-X_i(t)\right ),  \label{eq:rxt_defn}   
\end{align}
as
\begin{align}
  \msc C( X,t;Y,t') = \langle \hat \varrho(X,t)\hat \varrho(Y,t') \rangle_c =\langle \hat \varrho(X,t)\hat \varrho(Y,t') \rangle - \langle \hat \varrho(X,t)\rangle \langle\hat \varrho(Y,t') \rangle,
  \label{eq:rxt_ryt'_defn}
\end{align}
where  the subscript `c' represents the cumulant, and the averages $\langle ... \rangle $ are performed over initial configurations. The correlation at macroscopic scale is defined in terms of coarse-grained density $\varrho(\mt X,\mtt)$, defined as mesoscopic mean of the microscopic density
\begin{align}
\varrho(\mt X,\mtt) = \frac{1}{\Delta X}\int_{\ell \mt X-\Delta X/2}^{\ell \mt X+\Delta X/2} dZ ~\hat{\varrho}(Z,t=\ell \mt t),
\label{def:rho_cg}
\end{align}
where $a \ll \Delta X  \ll \ell$ is the coarse-graining length scale. The macro-scale correlation is  defined as 
\begin{align}
\mc C(\ell \mt X,\ell \mtt;\ell \mt Y,\ell \mtt')=\langle  \varrho(\mt X,\mtt) \varrho(\mt Y,\mtt') \rangle_c =\langle  \varrho(\mt X,\mtt) \varrho(\mt Y,\mtt') \rangle - \langle  \varrho(\mt X,\mtt)\rangle \langle \varrho(\mt Y,\mtt') \rangle.
\label{def:hd_corr}
\end{align}
By mapping to point particle gas and performing microscopic calculation, we find an exact analytical expression for the microscopic correlation $\msc C(X,t;Y,t')$. On the other hand, for the macro-scale correlation  $\mc C(\ell \mt X,\ell \mtt;\ell \mt Y,\ell \mtt')$ we use BMFT and find an explicit analytical expression. Both analytical results are verified through direct numerical simulation of the microscopic dynamics. We show that, as expected, while $\msc C(X,t;Y;t')$ displays correlation structures at microscopic length scale $a$, the correlation $\mc C(\ell \mt X,\ell \mtt;\ell \mt Y,\ell \mtt')$ exhibits behavior only at large space time scale $\mt X=X/\ell$ and $\mtt =t/\ell$.

As mentioned previously, the BMFT is based on two assumptions: hydrodynamic projections and the principle of local relaxation \cite{doyon2023ballistic}. Since these principles are not derived rigorously, it is important to verify the predictions of this theory from microscopic computations. We do this by observing that due to the definition in Eq.~\eqref{def:rho_cg}, the macro-scale correlation $\mc C$ is indeed a coarse-grained version of the micro-scale correlation $\msc C$ as 
\begin{align}
\mc C(\ell \mt X,\ell\mtt;\ell\mt Y,\ell\mtt') =
\frac{1}{\Delta X^2}\int_{\ell \mt X-\Delta X/2}^{\ell \mt X+\Delta X/2} dZ \int_{\ell \mt Y-\Delta X/2}^{\ell \mt Y+\Delta X/2} dZ'~ \msc C(Z,t;Z',t').
\label{rel:corr-macro-micro}
\end{align}
By coarse-graining the micro-scale correlation $\msc C(X,t; Y,t')$ data, we computed the macro-scale correlation and compared it with the prediction from BMFT to validate the hydrodynamic assumptions in this theory.

\section{Average mass density profile}
\label{sec:den}
 We now briefly review the microscopic evaluation of the average mass density profile $\bar \varrho(X,t)= \langle \hat{\varrho}(X,t)\rangle$ of hard rods at arbitrary time $t,$ as recently obtained in \cite{Mrinal_2024_HR}. 
Taking  average  over the initial configuration $\{X_i,v_i\}$ sampled from the ensemble (described in the previous section), one can compute the mean mass density $\bar \varrho(X,t)$. This computation can be easily done by mapping to an equivalent problem in HPG. 

As shown previously \cite{percus1969exact, Mrinal_2024_HR, singh2024thermalization}, each configuration of $N$ hard rods can be mapped to a unique configuration of $N$ hard point particles that evolve ballistically as non-interacting particles. Consequently, the dynamics of $N$ hard rods can be solved by first solving it in the point particle picture and then mapping it back to the hard rod coordinates. Using this procedure one finds that the mean mass density of hard-rods is given by \cite{Mrinal_2024_HR}
\begin{align}\label{eq:rho_Xt_micro}
    \bar {\varrho}(X,t)=\langle \hat{\varrho}(X,t)\rangle=\sum_{i=1}^N N \binom{N-1}{i-1}~ \big[q(x_i,t)\big]^{i-1}~ p(x_i,t)~ \big[1-q(x_i,t)\big]^{N-i},
\end{align}
where $x_i =X-(i-1)a$ and $p(x,t)$ represents the probability of finding a point particle at position $x$ at time $t$. This probability is given explicitly by 
\begin{align}\label{eq:prob_p}
 p(x,t)=\int^{\infty}_{-\infty} dy~g(x,t|y,0)~ \phi (y),
\end{align}
where $g(x,t|y,0) = \int dv ~\delta(x-y -vt) h(v) =\frac{1}{t} h\left(\frac{x-y}{t} \right)$ represents the propagator of a point particle to reach $x$ at time $t$ starting from $y$. The quantity 
\begin{align}\label{eq:prob_q}
      q(x,t) = \int^{x}_{-\infty} p(z,t) ~dz,
\end{align}
corresponds to the probability of finding a particle to the left of position $x$ at time $t$. 

\begin{figure}[t]
\centering
\includegraphics[width=9cm, height=6.5cm]{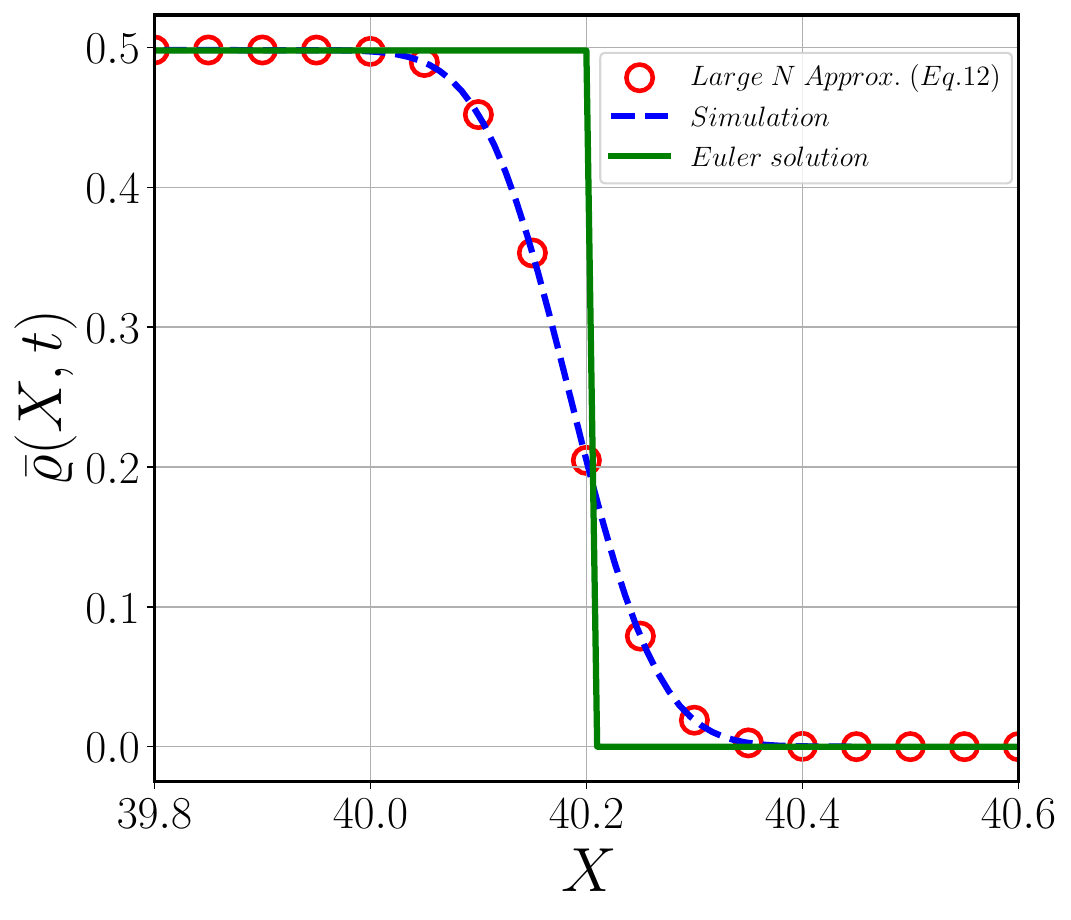}
\vspace*{-.2 cm}
\caption{Comparison of the density profiles of the special component obtained from the expression in Eq.~\eqref{sol:beyond-Euler-2}, and numerical simulations at $t=40,$ for $\hat{\varrho}_0=0.5.$ Here, the blue dashed line denotes simulation data, red circles represent the analytical solution [Eq.~\eqref{sol:beyond-Euler-2}], and green solid line corresponds to the Euler GHD solution \cite{Mrinal_2024_HR,singh2024thermalization} The profile is shown near the shock region, where the large $N$ analytical result exhibits excellent agreement with the numerical data, compared to the Euler GHD solution. The other parameters are $N=1000, a=0.01, v_0=1$, and averages are taken over $10^5$ realizations.   }
\label{Fig:largeN_compare}
\end{figure}

The expression for the microscopic mass density profile in Eq.~\eqref{eq:rho_Xt_micro} can be approximated in the thermodynamic limit, by taking large $N$ and small $a$, while keeping $\frac{Na}{\sigma}$ fixed. The approximate form of $\bar{\varrho}(X,t)$ is given by \cite{Mrinal_2024_HR} 
\begin{align}
\bar {\varrho}(X,t) &\approx \int_{-\infty}^\infty dz \frac{\sqrt{N}\exp \left( - \frac{N(X-z-a\bar F(z,t))^2}{2 a^2\Sigma_a^2(z,t)}~ \right)}{\sqrt{2 \pi a^2 \Sigma_a^2(z,t)}}~\bar \rho(z,t), 
\label{sol:beyond-Euler-2} 
\end{align}
where, 
\begin{align}
 \bar \rho(z,t) =Np(z,t),~\text{and}~\bar F(z,t)=\int_{-\infty}^zdy~\bar \rho(y,t)=Nq(z,t),
\end{align} 
are the mass density and cumulative mass density in the point particle representation, respectively and $\Sigma_a^2(z,t) = \bar F(z,t) (N - \bar F(z,t))$.  In the $N \to \infty$ limit (keeping $a$ fixed), the expression in Eq.~\eqref{sol:beyond-Euler-2} correctly reduces to the Euler solution 
\begin{align}
   \bar{\varrho}(X,t) = \int_{-\infty}^\infty dz~\delta(X-z-a\bar F(z,t))\bar \rho(z,t) =\frac{\bar \rho(z^*(X),t)}{1+a\bar \rho(z^*(X),t)}, 
   \label{sol:bar_varrho_Eu}
\end{align} 
where $z^*(X)$ is the solution of $X=z^* + a\int dz~\Theta(z^*-z)\bar\rho(z,t)$.
The expression in Eq.~\eqref{sol:beyond-Euler-2} 
 goes beyond the Euler GHD solution \cite{Mrinal_2024_HR,singh2024thermalization} for large but finite $N$.   It has been shown \cite{Mrinal_2024_HR,singh2024thermalization} that such deviations from the Euler solution can be seen better  for  the domain wall initial condition, in which an interface, initially at the origin, separates two species of hard rod gas, distinct by their velocity distributions, on the left and right side otherwise spatially uniformly distributed with density $\varrho_0$. In particular, we consider that all the rods on the left side have velocity $v_0$ (we call special rods) and the velocities of the rods on the right side are chosen from Maxwell distribution  with temperature $T=1$ (we call them background rods). Since the hard rod dynamics does not mix the momenta of different rods, the  total mass of the individual components remain unchanged, only their density profiles changes with time. Both the special and the background components have a jump in the density profiles at the origin at $t=0$. As time evolves, this jump moves ballistically according to the Euler solution. However, at diffusive space-time scale (as can be seen by  zooming at the jump) the domain wall also spreads. To describe such broadening,  one needs to find solutions beyond the Euler space-time scale, and that is given by the solution in Eq.~\eqref{sol:beyond-Euler-2}. We demonstrate this fact  in Fig. \ref{Fig:largeN_compare} where we compare  the density profile of the special component obtained in simulation and the solution in  Eq.~\eqref{sol:beyond-Euler-2}. We observe excellent agreement. 


In the next section \ref{sec:3}, we turn to the computation of space-time correlations of the mass density within the microscopic framework. We begin with the unequal space-time correlation, $\msc C(X,t;Y,0)$ in sec.~\ref{sec:C(t0)}, by examining how the density of a rod located at $X$ at time $t$ correlates with the mass density of another rod initially at position $Y.$ In section \ref{subsec:3_3}, we analyze the equal-time correlation $\msc C(X,t;Y,t)$, which characterizes how, at a given time $t,$ the mass density of a hard rod at position $X$ is correlated with that at $Y.$ 


\section{Correlation of the mass density: microscopic approach} \label{sec:3}
In this section, we focus on the main purpose of the paper, {\it i.e.} to find the space-time correlation for the mass density of hard rods following a microscopic approach. We first look at the space time correlation of the mass density, defined in Eq.~\eqref{eq:rxt_ryt'_defn}
with
\begin{align}\label{eq:rXtrYt'_def}
    \langle \hat{\varrho}(X,t) \hat{\varrho}(Y,t')\rangle=\sum_{i=1}^N \sum_{j=1}^N\langle \delta(X-X_i(t)) ~ \delta(Y-X_j(t')) \rangle, 
\end{align}
where $\langle \delta(X-X_i(t))~  \delta(Y-X_j(t')) \rangle$, represents the joint probability density function of finding the $i^{\rm th}$ rod at $X$ at time $t$, and the $j^{\rm th}$ rod at $Y$ at time $t'.$ In the point particle representation, such hard rod configurations correspond to configurations in which the $i^{\rm th}$ point particle  is at $x_i=X-(i-1)a$ at time $t$ and the $j^{\rm th}$ point particle is at $y_j=Y-(j-1)a$ at time $t'$. This means there are $(i-1)$ point particles below $x_i$ at time $t$ and $(j-1)$ point particles below $y_j$ at time $t'$. Denoting the probability of such configuration by $\msc P_{ij}(x_i,t;y_j,t')$ we rewrite the space time correlation as 
\begin{align}\label{eq:rXtrYt'_def_pp_full}
    \langle \hat{\varrho}(X,t) \hat{\varrho}(Y,t')\rangle=  \sum_{i=1}^N \sum_{j=1}^N \mathscr{P}_{ij}(x_i,t;y_j,t').
\end{align}
The average density $\langle \hat \varrho(X,t) \rangle$  was computed in the previous section. Subtracting the product $\langle \hat\varrho(X,t) \rangle \langle \hat\varrho(Y,t') \rangle$ from  $\langle \hat\varrho(X,t)  \hat\varrho(Y,t') \rangle$ yields the connected correlation $\msc C( X,t;Y,t')$, as defined in Eq.~\eqref{eq:rxt_ryt'_defn}. Hence, we next  focus on computing the joint probability density $\langle \hat{\varrho}(X,t) \hat{\varrho}(Y,t')\rangle$ which is performed separately for $t'=0, t>0$  and $t'=t>0$  in the next two sections.
The space-time correlation of the coarse-grained mass density has recently been investigated in \cite{AnupamMFT} using ballistic macroscopic fluctuation theory, where exact analytical results were obtained for both generic non-equilibrium and homogeneous equilibrium initial conditions. Recall we aim to compute this correlation directly from the microscopic dynamics and understand its relation with the hydrodynamic scale correlation.  

\subsection{Space-time correlation $\msc C(X,t;Y,0)=\langle \hat{\varrho}(X,t) \hat{\varrho}(Y,0)\rangle_c$}
\label{sec:C(t0)}
We start by evaluating the joint probability $ \mathscr{P}_{ij}(x,t;y,0)$ that the $i^{\rm th}$ point particle is at $x$ at time $t$ and the $j^{\rm th}$ point particle is at $y$ at $t'=0$.
To compute this probability
one needs to consider two distinct processes:: (i) the case where the $j^{\rm th}$ point particle starting from position $y$ evolves to the position $x$ at time $t$ and becomes the $i^{\rm th}$ particle; and (ii) a different particle (other than the $j^{\rm th}$) becomes the $i^{\rm th}$ particle at time $t$ and reaches at $x,$ while the $j^{\rm th}$ particle starting from $y$ reaches a position either to the left or the right of the position $x$ at time $t$. Accounting for both contributions, one can write $\mathscr{P}_{ij}(x,t;y,0)$ as (see \ref{sec:appA} for derivation)
\begin{subequations}
\label{eq:conn_corr_jpdf_twotime-full}
\begin{align}
\label{eq:conn_corr_jpdf_twotime}
        \mathscr{P}_{ij}(x,t;y,0) = & ~N \binom{N-1}{j-1}  q(y,0)^{j-1} p(y,0) \left[1-q(y,0) \right]^{N-j} 
        \\
        &\times\binom{N-1}{i-1} q(x,t)^{i-1} \left[ 1-q(x,t)\right]^{N-i} 
        \left[ g(x,t|y,0) + p(x,t) \mathbb{S}_{ij}(x,t;y,0) \right],
        \notag 
\end{align}
where
\begin{align}
    \mathbb{S}_{ij}(x,t;y,0)= \frac{(i-1) g_<(x,t;y,0)}{q(x,t)} +  \frac{(N-i) g_>(x,t;y,0)}{1-q(x,t)}. 
\end{align}  
\end{subequations}
Recall that $g(x,t|y,0)$ represents the propagator of a point particle to reach $x$ at time $t$ starting from $y$, and $p(x,t), q(x,t)$ are given in Eqs.~\eqref{eq:prob_p}-\eqref{eq:prob_q}. Here
\begin{align}\label{eq:prop-sp-ballistic_grl}
    g_<(x,t;y,0)=\int_{-\infty}^{x}~dz~g(z,t|y,0),~~
    g_>(x,t;y,0)=\int_{x}^{\infty}~dz~g(z,t|y,0),
\end{align}
denote the probabilities, in the point-particle representation, that a particle initially at position $y$ is found to the left or right of position $x$, respectively.
\begin{figure}[t]
\centering
\includegraphics[width=15.6cm, height=5.6cm]{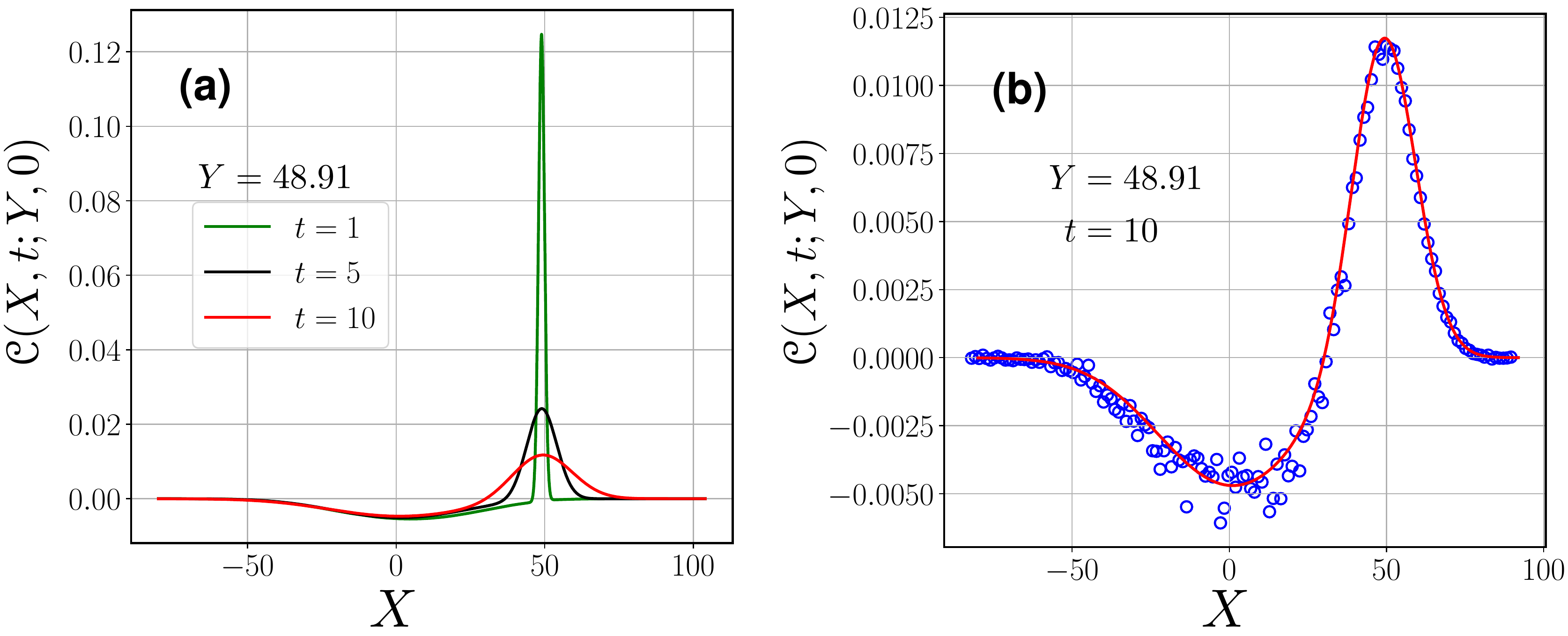}
\vspace*{-.2 cm}
\caption{
The plot (a) shows the evolution of the two-time density correlation $\msc C(X,t;Y,0)$ at $t=1,5,10$ for fixed $Y=48.91$ using the analytical exact expressions following Eq. \eqref{eq:conn_corr_two_time}. The plot (b) shows the comparison of the numerical simulation (blue circles) with the analytical exact result (red line) for $t=10$. 
using the mapping of Eq.~\eqref{eq:HR to HP mapping infinite line}. 
The parameters considered here are $N=200,~a=0.02,~\sigma=20$, and $T=1$. For numerical results, the averages have been taken over $5 \times 10^{10}$ initial configurations.} 
\label{Fig:t0_10_corr_plot}
\end{figure}
The first line on the right hand side of Eq.~\eqref{eq:conn_corr_jpdf_twotime} corresponds to the probability that the $j^{\rm th}$ point particle is at $y$ at $t'=0$. Given this event, the second line computes the probability that the $i^{\rm th}$ particle is at $x$ at time $t$. Of course this gets contribution from the two processes (i) and (ii) mentioned above. The contribution from process (i) is accounted by the propagator $g(x,t|y,0)$ inside the square bracket on the second line of Eq.~\eqref{eq:conn_corr_jpdf_twotime}, whereas the other term denoted by $\mathbb S_{ij}$ represents the contribution from process (ii). 

Inserting the expression of $\mathscr{P}_{ij}(x,t;y,0)$ from Eq.~\eqref{eq:conn_corr_jpdf_twotime} into Eq.~\eqref{eq:rXtrYt'_def_pp_full} and using the expression of the averaged density $\langle \hat{\varrho}(X,t)\rangle$ given in Eq.~\eqref{eq:rho_Xt_micro}, we obtain the following explicit expression for the microscopic space-time correlation
  \begin{align}\label{eq:conn_corr_two_time}
  \begin{split}
    \msc C(X,t;Y,0)
    =&  N \sum_{i=1}^N \sum_{j=1}^N \binom{N-1}{j-1}   q(y_j,0)^{j-1} p(y_j,0) \left[1-q(y_j,0) \right]^{N-j}    \\
       & ~\times~ \binom{N-1}{i-1}~q(x_i,t)^{i-1} \left[ 1-q(x_i,t)\right]^{N-i} \\ 
       &~\times~\left[ g(x_i,t|y_j,0) + p(x_i,t) \left( \mathbb{S}_{ij}(x_i,t;y_j,0) -N \right) \right],
\end{split}
\end{align}
where $x_i= X-(i-1)a$ and $y_j= Y-(j-1)a$.


In Fig. \ref{Fig:t0_10_corr_plot}, we compare this expression with the results from numerical simulations. For this plot, the initial configuration of the rods are chosen in the point-particle picture (see discussion after Eq.~\eqref{eq:prob_init}) with the distribution
\begin{align}\label{eq:init_corr_t0}
    \phi(x)=\frac{1}{\sqrt{2\pi\sigma^2}}~\exp\Big(-\frac{x^2}{2\sigma^2}\Big),~~h(v)=\frac{1}{\sqrt{2\pi T}}~\exp\Big(-\frac{v^2}{2T}\Big),
\end{align}
and then mapped to the hard rod configuration using Eq.~\eqref{eq:HR to HP mapping infinite line}. For all the plots from this section onward, we choose this initial condition.
Fig. \ref{Fig:t0_10_corr_plot} (a) shows how the space-time density correlation evolves with time using the exact analytical result, represented at times $t=1,5,10$. As time increases, the initially localized correlation around $X=Y$ starts broadening, while remaining centered around $Y$, indicating the spreading of density correlation at finite times. The times considered here are larger than microscopic collision times to probe the finite-time dynamics effect in density correlations. In Fig. \ref{Fig:t0_10_corr_plot} (b), the analytical exact result in Eq.~\eqref{eq:conn_corr_two_time} is compared with the numerical simulations at $t=10$, showing a good agreement. 

\subsection{Equal time correlation  $\msc C(X,t;Y,t)=\langle \hat{\varrho}(X,t) \hat{\varrho}(Y,t)\rangle_c$:}\label{subsec:3_3}
For equal times $t=t'$, in order to compute the connected correlation $\msc C(X,t;Y,t)$, we first evaluate the joint density $\langle \hat{\varrho}(X,t) \hat{\varrho}(Y,t)\rangle$ defined as
\begin{align}
    \langle \hat{\varrho}(X,t) \hat{\varrho}(Y,t)\rangle=\sum_{i=1}^N \sum_{j=1}^N\langle \delta(X-X_i(t)) ~ \delta(Y-X_j(t)) \rangle.
\end{align}
Naturally, this quantity takes a non-zero value only for the separation satisfying $|X-Y| \ge a$. Therefore for $|X-Y| \ge a$ we can write 
\begin{align}\label{rxryt_expr1}
    \langle \hat{\varrho}(X,t) \hat{\varrho}(Y,t)\rangle=\Bigg\langle \sum_{i\neq j=1}^N  \delta(X-X_i(t)) ~ \delta(Y-X_j(t)) \Bigg\rangle.
\end{align}
One can show that $\langle \hat{\varrho}(X,t) \hat{\varrho}(Y,t)\rangle$ can be expressed as (see \ref{sec:appB} for derivation)  
\begin{subequations}\label{eq:rxryt_expr_full}
\begin{align}\label{eq:rxryt_expr}
    \langle \hat{\varrho}(X,t) \hat{\varrho}(Y,t)\rangle=N (N-1) \sum_{m=0}^{N-2} \sum_{k=0}^{\tilde{k}(X,Y,m)} \mathcal{P} (\max(X,Y),\min(X,Y),m,k,t;N),
\end{align}
\begin{align}
   {\rm where}~~ \tilde{k}(X,Y,m) = \min \Bigg[ \left\lfloor \frac{|X-Y|-a}{a} \right\rfloor ,N-m-2 \Bigg],
\end{align}
and
\begin{align}\label{eq:jont_dist_same_t_two_point}
    \mathcal{P}(X,Y,m,k,t;N) =& \binom{N-2}{m} \binom{N-m-2}{k} \big[q(Y-ma,t) \big]^m ~p(Y-ma,t) \cr  & ~\times~\big[\bar q(X-(k+m+1)a,Y-ma,t) \big]^{k} ~p(X-(k+m+1)a,t) \cr &
    ~\times~\big[1-q(X-(k+m+1)a,t) \big]^{N-m-k-2}.
\end{align}
\end{subequations}
\begin{figure}[t]
\centering
\includegraphics[width=15.7cm, height=5.05cm]{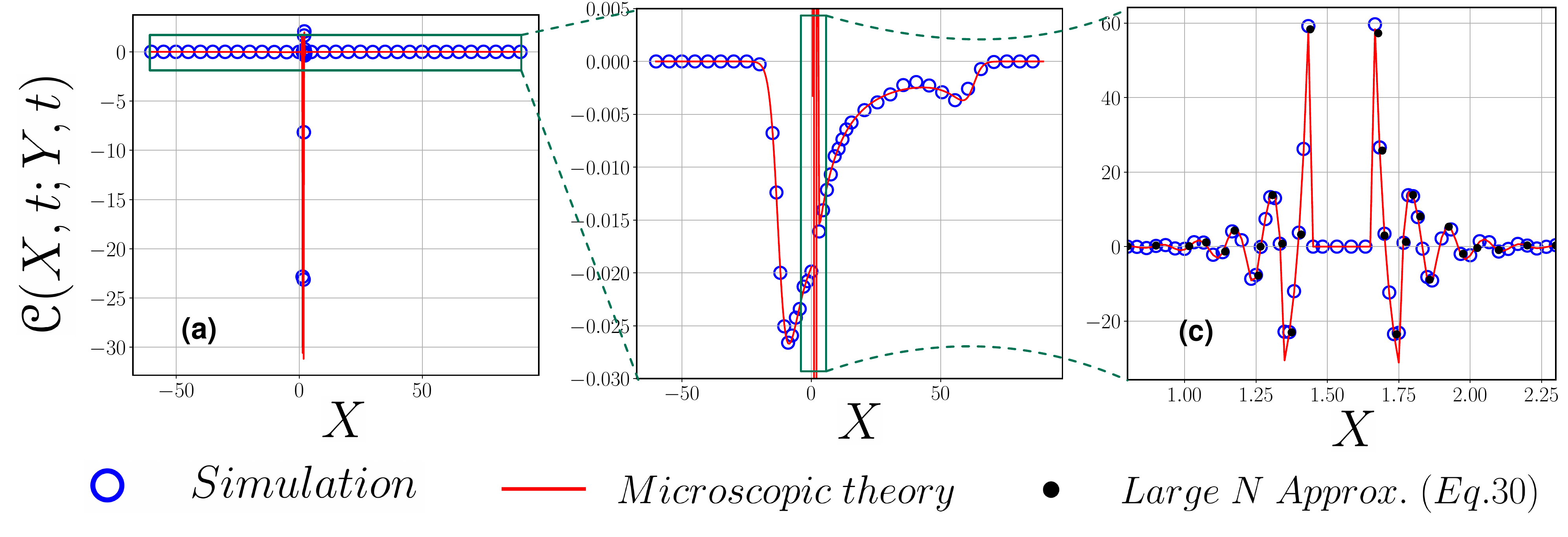}
\vspace*{-.2 cm}
\caption{Plots comparing the  mass density correlation $\msc C(X,t;Y,t)$ between the analytical exact expression (red solid lines) following Eq.~\eqref{eq:jont_dist_same_t_two_point} and the numerical simulation (blue circles) for $Y=1.55$ and $t=5$ at different levels of magnification.  Plot (a) shows the correlation over the full spatial range in $X$. Plot (b) presents a zoomed-in view around $X=Y$ position. To further resolve the local structure of the correlation around $X=Y$, plot (c) displays a finer zoom with more closely spaced data points. 
The values of other parameters considered for this plot are $N=500,a=0.1, \sigma=0.5,$ and $  T=1$ and $Y=1.55.$ The averages have been taken over $10^7$ initial configurations.}  
\label{Fig:samet_N_500_corr}
\end{figure}
Here, recall that the probabilities $p(x,t),q(x,t)$ in the point particle representation are defined in  
Eqs.~\eqref{eq:prob_p}-\eqref{eq:prob_q}, and
\begin{align}
    \bar q(a,b,t)=\int_{b}^{a}dz~p(z,t).
\end{align}
From the expression of $\langle \hat{\varrho}(X,t) \hat{\varrho}(Y,t)\rangle$ as in Eq.~\eqref{eq:rxryt_expr_full},  the exact microscopic form of the equal-time mass density correlation follows directly from Eq.~\eqref{eq:rxt_ryt'_defn}, where $\langle \hat{\varrho}(X,t)\rangle$ is given in Eq.~\eqref{eq:rho_Xt_micro}. Comparisons between the exact microscopic expressions of the correlation $\msc C( X,t;Y,t)$ of the mass density with numerical simulations are shown in Fig. \ref{Fig:samet_N_500_corr}. The simulations are performed for an initial configuration of rods prepared in the point particle representation with the distribution specified in Eq.~\eqref{eq:init_corr_t0}, subsequently mapped to the hard rod configuration using Eq.~\eqref{eq:HR to HP mapping infinite line}. Fig. \ref{Fig:samet_N_500_corr}(a) exhibits the unzoomed version of the correlation $\msc C(X,t;Y,t)$ for $Y=1.55$ and $t=5$ where we observe that this correlation indeed lives on microscopic scale. To see this better, we present zoomed-in view of this  plot in (b) and (c). While both (b) and (c), display excellent agreements between simulation data (symbols) and theory (solid lines), plot (c) reveals richer fluid like structures at microscopic length scale $a$.

To analyze the joint density $\langle \hat{\varrho}(X,t) \hat{\varrho}(Y,t)\rangle$ in the  large-$N$ limit, we work in the Fourier space. We introduce the Fourier transform of the joint density function as
\begin{align}\label{eq:rxryt_FFT}
\tilde {\mt P}(k_1,k_2,t)=\int_{-\infty}^{\infty}dX \int_{-\infty}^{\infty}dY~e^{\iota k_1X}e^{\iota k_2Y} \langle \hat{\varrho}(X,t) \hat{\varrho}(Y,t)\rangle,
\end{align}
such that the inverse transformation is
\begin{equation}\label{eq:rxryt_IFFT}
    \langle \hat{\varrho}(X,t) \hat{\varrho}(Y,t)\rangle=\int_{-\infty}^{\infty}\frac{dk_1}{2\pi} \int_{-\infty}^{\infty}\frac{dk_2}{2\pi}~e^{-\iota k_1X}e^{-\iota k_2Y}~\tilde {\mt P}(k_1,k_2,t).
\end{equation}
Substituting the expression of $\langle \hat{\varrho}(X,t) \hat{\varrho}(Y,t)\rangle$ from Eq.~\eqref{eq:rxryt_expr_full} into Eq.~\eqref{eq:rxryt_FFT} and simplifying yields
\begin{align}
    \tilde {\mt P} (k_1,k_2,t)&= \int_{-\infty}^{\infty}dz_1 \int_{-\infty}^{\infty}dz_2~e^{\iota k_1z_1}e^{\iota k_2z_2}~\bar \rho(z_1, t)~\bar \rho(z_2-a,t)\cr
     &~~\times ~\left[1+\left(e^{\iota (k_1+k_2)a}-e^{ik_2a}\right)\frac{\bar F(z_1,t)}{N}+\left(e^{\iota k_2a}-1\right)\frac{\bar F(z_2-a,t)}{N}\right]^{N-2},\label{eq:largeNrxry1}
\end{align}
where recall $\bar \rho(z,t)=Np(z,t)$ be the mass density and $\bar F(z,t)=Nq(z,t)=\int_{-\infty}^z dy~\bar \rho(y,t)$ be the cumulative mass density in the point particle representation. The Eq.~\eqref{eq:largeNrxry1} remains exact for any finite values of $N$ and $a$. To obtain the corresponding distribution function in the Euler scale, we first take $N \to \infty$ limit and then take small $\frac{a}{\sigma}$ limit. In this limit, with the approximation of $e^{\iota ka}=1+\iota ka+{\cal O}(k^2a^2),$ we arrive at the form 
\begin{align}\label{eq:rhryt_euler1}
  \langle \hat{\varrho}(X,t) \hat{\varrho}(Y,t)\rangle  =\frac{\bar \rho\Big(z_1^*(X),t\Big)}{1+a\bar \rho\Big(z_1^*(X),t\Big)}~\times ~ \frac{\bar \rho\Big(z_2^*(Y)-a,t\Big)}{1+a\bar \rho\Big(z_2^*(Y)-a,t\Big)},
\end{align}
where $z_1^*(X)$ and $z_2^*(Y)$ are the solutions of the equations
\begin{equation}\label{eq:rhryt_euler2}
    X=z_1^*+a\bar F(z_1^*,t),~Y=z_2^*+a\bar F(z_2^*-a,t).
\end{equation}
The Eqs.~\eqref{eq:rhryt_euler1}-\eqref{eq:rhryt_euler2} provide the Euler GHD solution of the joint density $\langle \hat{\varrho}(X,t) \hat{\varrho}(Y,t)\rangle.$ Now, to obtain an approximate expression for the joint density function at large but finite $N$, we again rewrite the $[\dots]^{(N-2)}$ term in Eq.~\eqref{eq:largeNrxry1} as $\exp\Bigg[(N-2) \log \Big[1+\left(e^{\iota (k_1+k_2)a}-e^{\iota k_2a}\right)\frac{\bar F(z_1,t)}{N}+\left(e^{\iota k_2a}-1\right)\frac{\bar F(z_2-a,t)}{N}\Big]\Bigg]$ and then expand in powers of $1/N$ and $a,$ considering the regime of large $N$ and small $a$  keeping $\frac{Na}{\sigma}$ finite. This leads to the expression of $\langle \hat{\varrho}(X,t) \hat{\varrho}(Y,t)\rangle$ as (see \ref{sec:appB1} for derivation)
\begin{align}\label{eq:largeNrxryt_1}
    \langle \hat{\varrho}(X,t) \hat{\varrho}(Y,t)\rangle = \frac{1}{(2\pi)^2}\int_{-\infty}^\infty dz_1 \int_{-\infty}^\infty dz_2 ~ \bar \rho(z_1, t) \bar \rho(z_2-a, t) \frac{1}{ \sqrt{\det(\mathbf A)}} \exp\left(-\frac{1}{2} \mathbf{Y}^T \mathbf A^{-1} \mathbf{Y}\right)
\end{align}
with
\begin{equation}
    \mathbf{Y} = \begin{pmatrix} X - \mu_1 \\ Y - \mu_2\end{pmatrix}, \hspace{0.5cm} \mu_1= z_1 + a \bar F(z_1, t), \hspace{0.3cm} \mu_2=z_2 + a \bar F(z_2-a, t),
\end{equation}
\begin{gather}
   {\rm and}~~ \mathbf{A} = \frac{a^2}{N} \begin{pmatrix} \bar F(z_1, t) \big( N - \bar F(z_1, t) \big) & \bar F(z_1, t) \big( N - \bar F(z_2 - a, t) \big) \\
\bar F(z_1, t) \big( N - \bar F(z_2 - a, t) \big) & \bar F(z_2 - a, t) \big( N - \bar F(z_2 - a, t) \big) \end{pmatrix}.
\end{gather}
The Eq.~\eqref{eq:largeNrxryt_1} provides the macroscopic distribution function of the joint distribution $\langle \hat{\varrho}(X,t) \hat{\varrho}(Y,t)\rangle$, which extends beyond the Euler space-time scale. In this large $N$ limit, the corresponding expression for  $\langle \hat{\varrho}(X,t)\rangle$ is provided in Eq.~\eqref{sol:beyond-Euler-2}. Substituting the expressions of $\langle \varrho(X,t)\rangle$ and $\langle \hat{\varrho}(X,t) \hat{\varrho}(Y,t)\rangle$ from Eqs.~\eqref{sol:beyond-Euler-2} and \eqref{eq:largeNrxryt_1} one obtains the equal-time correlation for the mass density in the large $N$ limit following Eq.~\eqref{eq:rxt_ryt'_defn}. In Fig. \ref{Fig:samet_N_500_corr}(c), we further compare the correlation $\msc C(X,t;Y,t)=\langle \hat{\varrho}(X,t) \hat{\varrho}(Y,t)\rangle_c$ obtained using the large $N$ approximation analysis (discs) with the exact microscopic expression (solid line) and with the numerical simulation (circles). The large $N$ approximation exhibits excellent quantitative agreement with both exact expression in Eq.~\eqref{eq:rxryt_expr_full} and simulation results.

\section{Hydrodynamic scale correlations of the mass density} \label{sec:4}
The hydrodynamic description of the hard-rod system is formulated in terms of the phase-space density, which evolves according to the Euler scale generalized hydrodynamics (GHD) equation in the ballistic space-time scale \cite{Mrinal_2024_HR,singh2024thermalization}. To establish a hydrodynamic description, one starts with an initial ensemble characterized by phase space density varying slowly over a large length scale $\ell$. In our context (see after Eq.~\eqref{eq:HR to HP mapping infinite line}), such an initial state can be achieved by choosing the mass density function $\bar \rho(x)=N\phi_0(x/\ell)$ while keeping the velocity distribution  $h(v)$ same as before. The large scale motion of the system can be described by the coarse-grained phase space density $\mf f(\mt X,v,\mtt)$ defined as 
\begin{align}
\begin{split}
    \mf {f}(\mt X,v,\mt t)&= \frac{1}{\Delta X}\int_{\ell \mt X-\Delta X/2}^{\ell \mt X+\Delta X/2} dZ ~\hat {\mf f}(Z,v,t=\ell \mt t), \\ 
    \text{where}~\hat {\mf f}&(Z,v,t) = \sum_{i=1}^N\delta(Z-X_i(t))\delta(v-v_i),
    \end{split}
    \label{def:mfb_f}
\end{align}
is the empirical phase space density. The coarse-graining scale $\Delta X$ is chosen so that $a \ll \Delta X \ll \ell$. One can consider $\Delta X \sim \ell^\nu$ with $0<\nu<1/2$ \cite{hubner2025hydrodynamics}. For large $\ell$, it is expected   that the density $\mf{f}(\mt X,v,\mt t)$, starting from a profile $\mf f(\mt X,v,0)$, evolves according to the Euler GHD equation \cite{hubner2025hydrodynamics, doyon2023ballistic}
\begin{align}
\partial_{\mt t} \mf f(\mt X,v,\mt t)+\partial_{\mt X}\big(v_{\rm eff}(v) \mf f(\mt X,v,\mt t)\big)=0, \label{eq:eghd}
\end{align}
on ballistic space-time scale ($\mt X \sim \mt t$), where the effective velocity is 
\begin{align}
v_{\rm eff}(v)&= 
\frac{v-a\int dv~v\mf f(\mt X,v,\mt t)}{1-a \varrho(\mt X,\mt t)},
\end{align}
and the coarse grained mass density is 
$\varrho(\mt X,\mtt) = \int dv ~\mfb{f}(\mt X,v, \mt t)$ which essentially is same as in Eq.~\eqref{def:rho_cg}.
We are interested in computing the  space-time correlation $\mc C$ of coarse grained mass density defined in Eq.~\eqref{def:hd_corr}.
For each microscopic  configuration $\{X_i,v_i\}$ of the rods chosen from the initial ensemble, one can construct the coarse-grained density $\mf f(\mt X,v,0)$ which is clearly fluctuating. It can be shown that \cite{AnupamMFT}, the probability of  $\mf f(\mtX,v,0)$ can be described by a large deviation form characterized by an appropriate Free energy functional $\mc F^{\rm r}[\mf f(\mt Y,v,0)]$
  \begin{align}
    \mc{P}_{\rm r}[\mf f(\mtX,v,0)] &\asymp  e^{-\ell \mc{F}^{\rm r}[\mf f(\mt X,v,0)]}~\delta\left( \ell\int d\mt X \int dv~ \mf f(\mt X,v,0) - N\right), 
    \label{mcP[mff(0)]}
\end{align}
where the delta function ensures the mass normalization. 

As the rods move under the microscopic collision dynamics, the fluctuating phase space density evolves and generates space-time correlation. Since, by definition the above hydro-scale correlation is related to the microscopic mass density correlation $\msc C(Z,t;Z',t')=\left \langle \rho(Z,t) \rho(Z',t')\right \rangle_c$ as in Eq.~\eqref{rel:corr-macro-micro},
one can use it to compute the hydro-scale correlation. The micro-scale correlations are computed in Sec.~\ref{sec:3}. Another way to compute the hydro-scale correlation is BMFT \cite{doyon2023ballistic}.  In this section we compute, hydro-scale correlation $\mc C(X,t;Y,t')$ using both, microscopic theory and BMFT, and compare results from the two methods.

According to BMFT, the density fluctuations on Euler space-time scale are dominated by the initial density fluctuations that are coherently propagated to the current location by the Euler GHD Eq.~\eqref{eq:eghd}. Other possible noises that may dynamically emerge due to coarse graining procedure can be neglected. 
It has been shown in \cite{AnupamMFT} that the probability distribution of the time evolution history of $\mf f(\mtX,v,\mtt)$, in the Euler scaling limit, can be approximated by \cite{AnupamMFT}
\begin{align}
\mc{P}_{\rm r}[\mf f(\mt X,v,\mtt)] \asymp e^{- \ell \left(\mc{F}^{\rm r}[\mf f(\mt X,v,0)] -\mc{F}^{\rm r}[\mf f_{\rm eq}(\mt X,v,0)]\right)}~\delta\bigg[\partial_{\mtt} \mf f(\mt X,v,\mtt)+\partial_{\mt X}v_{\rm eff}[\mf f]\mf f(\mt X,v,\mtt)\bigg], \label{mcP[mff(t)]}
\end{align}
where $\mf f_{\rm eq}(\mt Y, u, 0)$ is the equilibrium profile that minimizes the free energy $\mc F^{\rm r}[\mf f(0)]$.
It is easy to see that the two-point correlation $\mc C(X_a=\ell \mtX,t_a=\ell \mtt_a;X_b=\ell \mt X_b,t_b=\ell \mtt_b)$ can be expressed as the following path integral 
\begin{subequations}
\begin{align}
\mc C(\ell\mt X_a, \ell\mtt_a;\ell\mt X_b, \ell\mtt_b) &= - \left[ \frac{\partial}{\partial \lambda}\frac{\left\langle \varrho(\mtX_b,\mtt_b)e^{-\lambda \varrho(\mtX_a,\mtt_a)}\right \rangle_{\mc P_{\rm r}[\mf f]}}{\left\langle e^{-\lambda \varrho(\mtX_a,\mtt_a)}\right \rangle_{\mc P_{\rm r}[\mf f]}} \right]_{\lambda=0}, \label{def:hd-corr-pathInt} \\ 
&=- \left[ \frac{\partial}{\partial \lambda}\left\langle \varrho(\mtX_b,\mtt_b) \right \rangle_{\mc P_{\rm r}^\lambda[\mf f]}\right]_{\lambda=0} 
,
\label{eq:hd-corr-pathInt} 
\end{align}
where $\mc P_{\rm r}^\lambda[\mf f]=\frac{\mc P_{\rm r}[\mf f]e^{-\lambda \varrho(\mt X_a,\mtt_a)}}{\left\langle e^{-\lambda \varrho(\mtX_a,\mtt_a)}\right \rangle_{\mc P_{\rm r}[\mf f]}}$ and for an observable $\mc O[\mf f(\mtt)]$,
\begin{align}
\left \langle  \mc O[\mf f(\mtt)] \right\rangle_{\mc P_{\rm r}^\lambda[\mf f]} = \int \mc D[\mf f(\mtX, v,\mtt)] \mc P_{\rm r}^\lambda[\mf f(\mt X,v,\mtt)]~\mc O[\mf f(\mt X,v,\mtt)].
\end{align}
\end{subequations}
In the thermodynamic limit, the path integral in Eq.~\eqref{eq:hd-corr-pathInt} can be performed by saddle point method (see \ref{sec:hrdro-corr} for details) and one finds
\begin{align}
  \bar\varrho_\lambda(\mt X_b,\mtt_b) = \langle \varrho (\mt X_b,\mtt_b)\rangle_{\mc P_{\rm r}^\lambda[\mf f]} \approx \varrho^{\rm sd}_\lambda(\mt X_b,\mtt_b), 
\end{align}
where $\varrho_\lambda^{\rm sd}(\mt X, \mtt)$ for $0 < \mtt < \mt T$ (with some $\mt T>\max \{\mtt_a,\mtt_b\} \ge 0$) is the saddle-point density profile that can be obtained by solving some appropriate saddle point equations. 
Consequently one finds that the correlation $\mc C$ has the following scaling form
\begin{align}\label{eq:corr_scaling}
    \mathcal{C}(X_a=\ell \mt X_a,t_a=\ell \mt t_a;& X_b=\ell \mt X_b,t_b=\ell \mt {t_b})=\frac{1}{\ell}~ \mt C(\mt X_a,\mt t_a; \mt X_b,\mt{t}_b), \\
    \text{with}~~&~\mt C (\mt X_a,\mt t_a; \mt X_b,\mt{t}_b) = -\left [\frac{\partial}{\partial \lambda} \varrho_\lambda^{\rm sd}(\mt X_b,\mtt_b)\right]_{\lambda = 0}.\label{eq:scaled-corr-rho_sd}
\end{align}
Recently, a similar saddle point approach has been used in the point particle picture to compute the correlation in generic classical integrable systems \cite{kethepalli2025ballistic}.
The details of the saddle point calculation and the computation of $\varrho_\lambda^{\rm sd}(\mt X, \mtt)$ are given in  \ref{sec:hrdro-corr}. Instead, we present only the results here.

\subsection{Equal time correlation $\mt C(\mt X_a,\mt t; \mt X_b,\mt{t})$:}
Solving the BMFT equations~\eqref{eq:saddle}, the equal time correlation of the mass density ({\it i.e.} at $\mt t_a =\mt t_b =\mt t$) at two distinct positions $\mt X_a$ and $\mt X_b$ is given by (see \ref{sec:appC} for derivation)
\begin{figure}[t]
\centering
\includegraphics[scale=0.3]{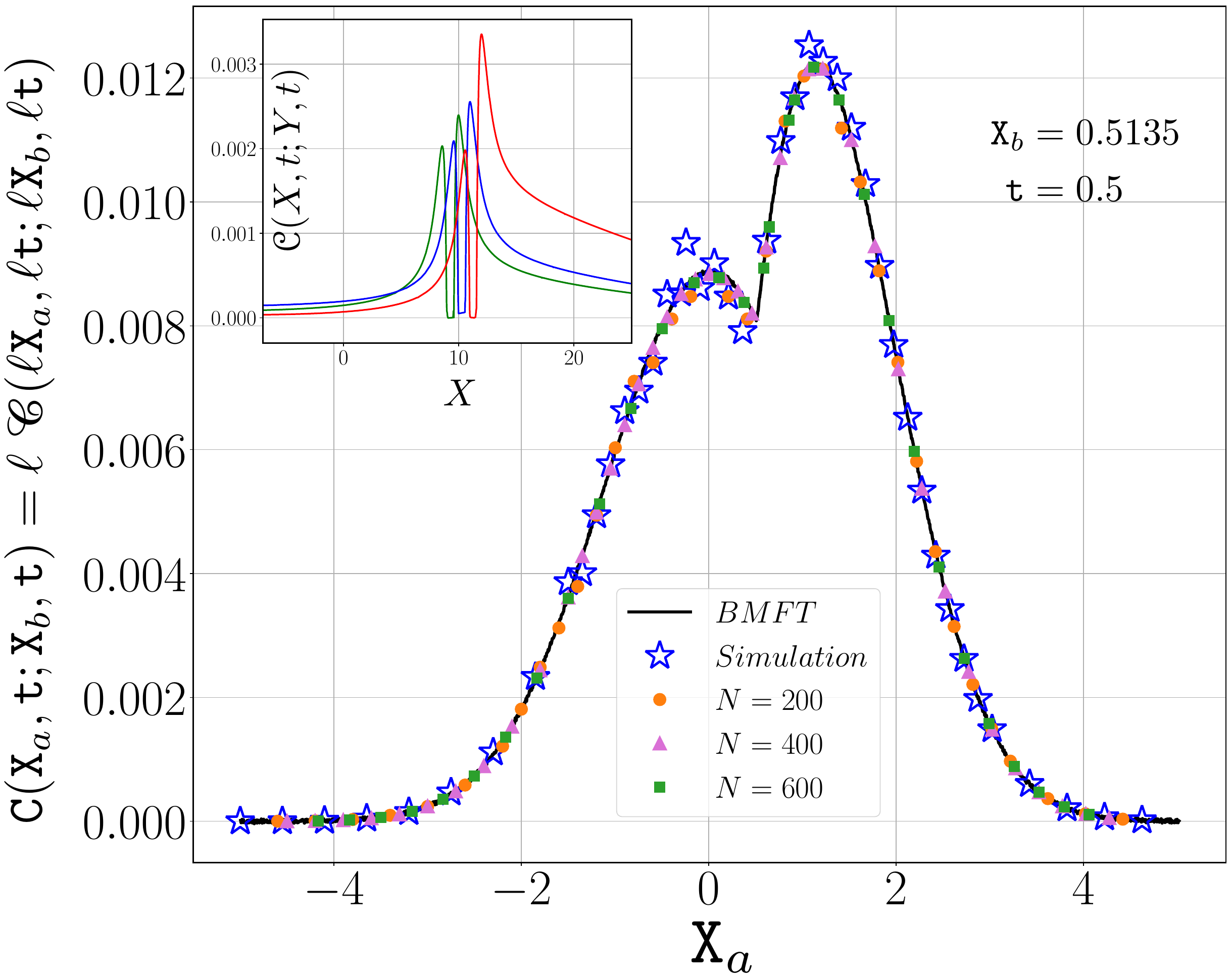}
\vspace*{-.2 cm}
\caption{ Plots compairing the equal-time connected correlation function derived from BMFT as in Eq.~\eqref{eq:corr_expr_bmft} [black solid line], with that obtained by coarse-graining  the analytical microscopic expression $\msc C(X,t;Y,t)$ (given in Eq.~\eqref{eq:rxt_ryt'_defn} using Eqs.~\eqref{eq:rho_Xt_micro}  and \eqref{eq:rxryt_expr_full}) [orange discs],  and the numerical simulation of the microscopic dynamics [blue star], for $N=200, a=0.002$. The coarse-graining has been performed for the macroscopic position  $\mt X_b=0.5135$ and $\mt t=0.5$ using fluid-cell averaging as described in Eq.~\eqref{def:rho_cg}, with the fluid cell size chosen as $\Delta X=0.1\ell,$ where $\ell=20.$  The coarse-grained equal-time correlation obtained from the microscopic dynamics shows very good agreement with the long-range correlations obtained theoretically by BMFT.  Additionally, we plot the coarse-grained correlation obtained from   microscopic correlation  for $N=400,600$ [triangle and square symbols] using the identical initial condition with $\ell=40$ and $60$ , respectively, for $a=0.002$, and at the same fixed macroscopic time $\mt t=0.5.$ The excellent collapse of the coarse-grained correlation functions for different $N$ and $\ell$ (with same ratio $\frac{\ell}{Na}$) verifies  the scaling relation in Eq.~\eqref{eq:corr_scaling}. {\bf Inset:} Plots of the corresponding microscopic correlation $\msc C(X,t;Y,t)$ for $N=200$ as functions of $X$ for few values of $Y$ fixed in the range $[ \mt X_b \ell- 0.1\ell,  \mt X_b \ell+ 0.1\ell]$ with $\mt X_b=0.5135$. These microscopic correlation functions are used on the right hand side of Eq.~\eqref{rel:corr-macro-micro} at the coarse-graining step  to obtain the hydro-scale correlation plotted in the main figure. The averages have been taken over $10^9$ initial configurations.
}
\label{Fig:t_cg_corr_scale}
\end{figure}
\begin{subequations}\label{eq:corr_expr_bmft_full}
\begin{align}
    \mt C(\mt X_a,\mt t; \mt X_b,\mt{t})=(1-a\bar{\varrho}(\mt X_a,\mt t))^2~\bar{\varrho}(\mt X_b,\mt t) ~\delta(\mt X_a - \mt X_b)+\mt C^{{\rm lr}}(\mt X_a,\mt t; \mt X_b,\mt{t}),
    \label{eq:mt_C-t-t}
\end{align}
where the long-range part of the correlation is expressed as
\begin{align}\label{eq:corr_expr_bmft}
    \mt C^{{\rm lr}}(\mt X_a,\mt t; \mt X_b,\mt{t})&=
   \mt{C}_{\rm r}(\mt X_a,\mt X_b,\mt t) \Theta(\mt X_b -\mt X_a) +  
    \mt{C}_{\rm r}(\mt X_b,\mt X_a,\mt t) \Theta(\mt X_a -\mt X_b),
\end{align}
with
\begin{align}
\mt{C}_{\rm r}(\mt X_a,\mt X_b,\mt t)=&\Big[a^2\big(\partial_{\mt 
 X_a}\bar{\varrho}(\mt X_a,\mt t)\big)\big(\partial_{\mt 
 X_b}\bar{\varrho}(\mt X_b,\mt t)\big) \bar{\mt F}(\mt X_a,\mt t) \notag \\ 
 &~~-a(1-a\bar{\varrho}(\mt X_a,\mt t))\bar \varrho(\mt X_a,\mt t)\big(\partial_{\mt 
 X_b}\bar{\varrho}(\mt X_b,\mt t)\big)\big].
 \end{align}
 \end{subequations}
Here  $\bar \varrho(\mt X, \mtt)$ represents the solution of the Euler GHD and  is given in Eq.~\eqref{sol:bar_varrho_Eu}, and $\bar{\mt F}(\mt X,\mt t)=\int_{-\infty}^{\mt X}d\mt Z~ \bar{\varrho}(\mt Z,\mt t)$ denotes the cumulative mass density. 
The correlation in Eq.~\eqref{eq:mt_C-t-t} has two parts, one singular part (containing $\delta$-function) and the other non-singular part. The singular part corresponds to the contribution from local-equilibrium where the smooth part represents the long-range part of the correlation. Such long-range correlations  are not observed in non-integrable systems and were first predicted in \cite{doyon2023emergence} for general integrable systems. The presence of such correlations are artifacts of the Euler evolution of inhomogeneous initial fluctuation \cite{doyon2023emergence,doyon2023ballistic} and the absence of dynamically emergent noise in integrable systems \cite{hubner2025hydrodynamics, doyon2025hydrodynamic}. The long-range part has been shown to have a crucial effect on the structure of the HD equation on the diffusion scale and leads to a  diffusive scale equation fundamentally different from the one obtained from the local-equilibrium approximation \cite{hubner2024diffusive}.

Returning to the context of the hard rod gas, we observe that for the homogeneous state $\bar {\mf f}(X,v) = \varrho_0 h(v)$, the long-range part in Eq.~\eqref{eq:corr_expr_bmft_full} vanishes and one recovers the short range correlation.
To test  the correspondence between hydrodynamic and microscopic descriptions given in Eq.~\eqref{rel:corr-macro-micro}, in Fig. \ref{Fig:t_cg_corr_scale}, we compare the above BMFT prediction  with the corresponding coarse-grained expression obtained from  the analytical microscopic expressions (in Eq.~\eqref{eq:rxryt_expr_full}) and the numerical simulation of the microscopic dynamics. 
For this analysis, the initial configuration of the rods is prepared in the point-particle representation with the distribution specified in Eq.~\eqref{eq:init_corr_t0}, and subsequently mapped to the hard rod configuration using Eq.~\eqref{eq:HR to HP mapping infinite line}. 
The excellent agreement implies that the coarse-grained equal-time correlation obtained from both the microscopic analytical expressions and the numerical simulations can predict the long-range correlations obtained using BMFT, and thereby validates the hydrodynamic assumptions used in BMFT. In Fig.~\ref {Fig:t_cg_corr_scale}, we further verify the scaling relation given in Eq.~\eqref{eq:corr_scaling} by plotting the scaled correlation curve for $N=400,600$ with $\ell=40$ and $60$, respectively, at the  macroscopic time $\mt t=0.5.$ The coarse-grained correlation functions for different $N$ and $\ell$ collapse onto a unique single curve, confirming the scaling form in Eq.~\eqref{eq:corr_scaling}.


\subsection{Space-time correlation $\mt C(\mt X_a,\mt t_a; \mt X_b,0)$:}
\begin{figure}[t]
\centering
\includegraphics[scale=0.35]{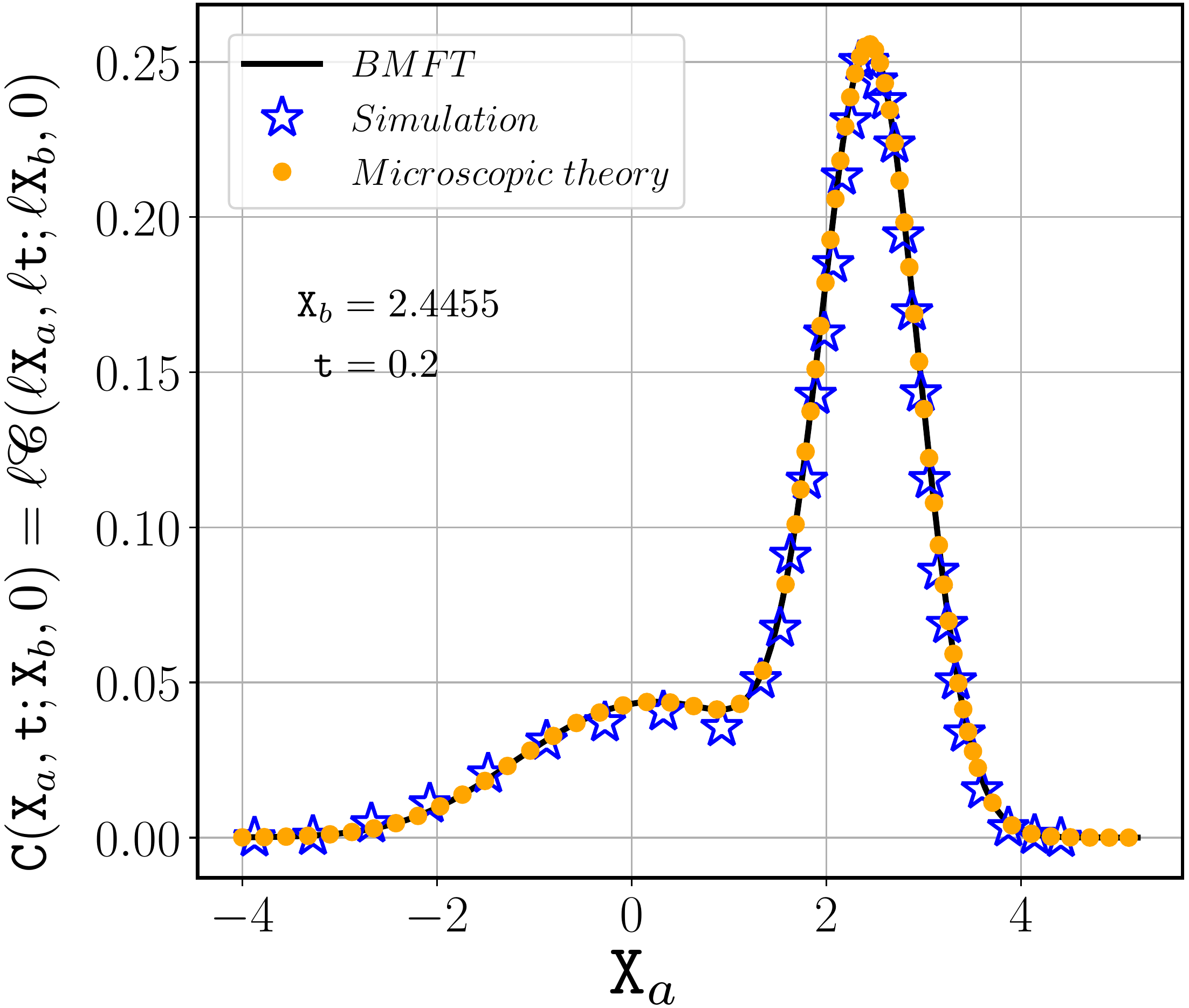}
\vspace*{-.2 cm}
\caption{ Comparison of the two-time connected correlation function obtained from BMFT (black solid line) as in Eq.~\eqref{eq:twotime_bmft_full}, with that derived by coarse-graining  the analytical microscopic expression in Eqs.~\eqref{eq:conn_corr_two_time} (orange discs) and the numerical simulation of the microscopic dynamics (blue stars). 
The coarse-graining has been performed for the macroscopic time $\mt t=0.2$ at the macroscopic position $\mt X_b=2.4455$ using fluid-cell averaging as described in Eq.~\eqref{def:rho_cg}, with the fluid cell size chosen as $\Delta X=0.08\ell,$ where $\ell=20.$ The coarse-grained equal-time correlation obtained from the microscopic dynamics shows an excellent agreement with the same obtained theoretically by BMFT. For this plot, we choose $N=200$ and $a=0.02$. The averages have been taken over $10^9$ initial configurations. 
}
\label{Fig:2t_cg_corr_bmft_sim_micro}
\end{figure}

Solving the saddle point equations in \eqref{eq:saddle} for $\mt t=0$, one can obtain the unequal space-time correlation as (see \ref{sec:appC} for derivation)
\begin{subequations}\label{eq:twotime_bmft_full}
    \begin{align}
        \mt C(\mt X_a,\mtt_a; \mt X_b,0)=\mt C_1(\mt X_a,\mtt_a; \mt X_b,0)+\mt C_2(\mt X_a,\mtt_a; \mt X_b,0),
    \end{align}
with
\begin{align}
  \mt C_1(\mt X_a,\mtt_a; \mt X_b,0)=&(1-a~\bar{\varrho}(\mt X_b,0))~\bar{\varrho}(\mt X_b,0) 
  \cr &~~\times
  ~\partial_{\mt X_a}
  \Bigg[(1-a~\bar{\varrho}(\mt X_a,\mt t_a))\mathcal{H}\Bigg(\frac{\mt x(\mt X_a)-\mt x(\mt X_b)}{\mt t_a}\Bigg)\Bigg],  
\end{align}
and
\begin{align}
    \mt C_2(\mt X_a,\mtt_a; \mt X_b,0)=&-a~ \partial_{\mt X_b}\bar{\varrho}(\mt X_b,0) \cr 
    &\times~\partial_{\mt X_a}
    \Bigg[(1-a~\bar{\varrho}(\mt X_a,\mt t_a))\int_{-\infty}^{\mt x(\mt X_b)}d\mt y~\bar{\varrho}(\mt y,0)\mathcal{H}\Bigg(\frac{\mt x(\mt X_a)-\mt y}{\mt t_a}\Bigg)\Bigg],
\end{align}
\end{subequations}
where $\mathcal{H}(u)=\int_{-\infty}^{u}dv~h(v)$ and $\mt x(\mt X)=\mt X-a\int_{-\infty}^{\mt X} d\mt Z~ \bar{\varrho}(\mt Z,\mt t)$. Here, $\bar \varrho(\mt X,\mtt)$ is the Euler solution of the mean mass density given in Eq.~\eqref{sol:bar_varrho_Eu} with initial condition $\bar{\varrho}(\mt X,0)$ provided in Eq.~\eqref{bar-varrho-ini}. The above expressions provide the space-time correlation of hard-rod mass density at the hydrodynamic scale.  Note that for $\bar {\mf f}(X,v) = \varrho_0 h(v)$, $\mt C(\mt X_a,\mtt_a; \mt X_b,0)$  in Eq.~\eqref{eq:twotime_bmft_full} reduces to the equilibrium space-time correlation previously obtained in Ref \cite{AnupamMFT}. In Figure \ref{Fig:2t_cg_corr_bmft_sim_micro}, we compare this result with microscopic correlations (from Eq.~\eqref{eq:conn_corr_two_time} and from numerical simulations) after explicit coarse-graining as in Eq.~\eqref{def:rho_cg} with $\Delta X=0.08\ell$ and   $\ell=20$. The good agreement once again verifies the micro-scale to hydro-scale correspondence in  Eq.~\eqref{rel:corr-macro-micro}. 


\section{Conclusion}\label{sec:5}

In this paper, we primarily compute the mass density correlations of a one-dimensional gas of hard rods at both microscopic and macroscopic scales. For a specific class of initial conditions, we perform exact analytical calculations of the microscopic correlation for both equal and unequal times. To compute correlation at macroscopic (hydrodynamic) scale, we define a coarse-grained mass density by taking the mesoscopic mean of the microscopic empirical density over a fluid cell. The fluctuations of this coarse-grained observable on the Euler space-time scale are assumed to be described using BMFT\cite{doyon2023ballistic, AnupamMFT}.

Using the BMFT approach, we derive an explicit expression for the coarse-grained mass density correlation, which reveals the emergence of long-range correlations on the hydrodynamic (Euler) scale \cite{doyon2023emergence}. Given that mesoscopic observables are derived from microscopic ones, a correspondence is expected between the hydro-scale and micro-scale correlations. Validating this micro-macro correspondence is essential for verifying the underlying assumptions of hydrodynamic theory. To date, such verifications have primarily been conducted numerically, focusing on mean density profiles \cite{chakraborti2021blast, ganapa2021blast, kumar2025shock, singh2024thermalization, Mrinal_2024_HR} and in certain cases the two point correlations in integrable systems \cite{doyon2023emergence}. By performing a systematic coarse-graining of our exact analytical expression of microscopic correlation, we compute the macro-scale correlation and demonstrate that it agrees precisely with the predictions of BMFT -- thus validating the assumptions of this theory in the context of one dimensional gas of hard rods.

Our present study can be extended to explore other observables of interest in hard rod systems. It would be worthwhile to compute, within the microscopic framework, the distribution of the net integrated current across a given location, as well as to analyze the fluctuations of this quantity. It would also be of interest to perform a parallel analysis of the fluctuations of the integrated current in the hard rod system using the recently developed BMFT for general initial conditions, and to examine whether the microscopic results can quantitatively reproduce the corresponding macroscopic predictions.

\section{Acknowledgments}
AK acknowledges the financial support under project ANRF/ARGM/2025/001207/MTR from the ANRF, DST, Government of India. IM, SC, and AK would also like to acknowledge the support from the DAE, Government of India, under Project No. RTI4001.

\appendix
\addtocontents{toc}{\fixappendix}
\section{Derivation of Eq.~\eqref{eq:conn_corr_jpdf_twotime-full}}
\label{sec:appA}
We first note that the joint probability density function $\mathscr{P}_{ij}(x,t;y,0)$ in the point particle picture is essentially
\begin{align}
    \mathscr{P}_{ij}(x,t;y,0)=\langle \delta(x-x_i(t)) ~ \delta(y-x_j(0)) \rangle
\end{align}
where the average is done over the (spatially) ordered configurations of the hard point particles with $i^{\rm th}$ particle at time $t$ is at $x$ and $j^{\rm th}$ particle initially at position $y$. This probability can be computed by considering two distinct possibilities: (i) the case where the $j^{\rm th}$ point particle starting from position $y$ evolves to the position $x$ at time $t$ and becomes the $i^{\rm th}$ particle; and (ii) a different particle (other than the $j^{\rm th}$) becomes the $i^{\rm th}$ particle at time $t$ and reaches at $x,$, while the $j^{\rm th}$ particle starting from $y$ reaches a position either to the left or the right of the position $x$ at time $t$. By summing these two contributions, we  write
\begin{align}
    \mathscr{P}_{ij}(x,t;y,0)=\mathcal{R}_{ij}(x, t; y,0)+\mathcal{S}_{ij}(x, t; y,0),
\end{align}
where $\mathcal {R}_{ij}$ and $\mathcal{S}_{ij}$ are the joint probability densities in the point particle representation corresponding to the processes (i) and (ii), respectively. To evaluate the contribution from process (i), we first consider a particle located at $y$ at $t=0$. The probability that it corresponds to the $j^{\rm th}$ particle is given by
\begin{align}
    P_{j} (y, 0)= N\binom{N-1}{j-1}~ \big[q(y,0)\big]^{j-1} ~p (y,0)~\big[1-q(y,0)\big]^{N-j},
\end{align}
where $p (y,0)$ denotes the probability of finding a point particle at position $y$ at $t=0$, $q(y,0)$ is the probability of finding a particle to the left of position $y$ at $t=0$ and recall their explicit expressions are given in Eqs.~\eqref{eq:prob_p}-\eqref{eq:prob_q}. $P_{j}(y,0)$ is then multiplied by the probability that it becomes the $i^{\rm th}$ particle at time $t.$ Thus, the probability that the $j^{\rm th}$ particle initially at $y$ directly evolves to the $i^{\rm th}$ particle at $x$ at time $t$ can be written as 
\begin{align}\label{eq:rxtry0_Rij}
    {\cal R}_{ij} (x, t;y, 0)&=\Bigg[\binom{N-1}{i-1} ~\big[q(x,t)\big]^{i-1}~ g(x,t|y,0) ~\big[1-q(x,t)\big]^{N-i}\Bigg]\cr
    &~~~~~\times \Bigg[N\binom{N-1}{j-1}~ \big[q(y,0)\big]^{j-1} ~p (y,0)~\big[1-q(y,0)\big]^{N-j}\Bigg],
\end{align}
where $g(x,t|y,0) = \int dv ~\delta(x-y -vt)  h(v) =\frac{1}{t} h\left(\frac{x-y}{t} \right)$ is the propagator that describes the transition of $j^{\rm th}$ point particle to reach at $x$ at time $t$ starting from $y$. 
The contribution from process (ii) is obtained as follows. We first evaluate the probability that a particle, initially chosen randomly at position $\bar x$ ($\bar x \neq y$), reaches the position $x$ at time $t$. This is then multiplied by the probability that the $j^{\rm th}$ particle is at $y$ at $t=0$ together with the propagator describing the transition of the $j^{\rm th}$ particle to $x'$ at time $t$ starting from $y$, where $x'\neq x$. Integrating over all $x'$ and treating separately the two cases $x'<x$ and $x'>x$ we obtain the joint probability density associated with this process as
\begin{align}\label{eq:rxtry0_Sij}
{\cal S}_{ij} (x, t;y, 0)&=  \Bigg[ (N-1)~\binom{N-2}{i-2} ~q(x,t)^{i-2}~p(x,t)~[1-q(x,t)]^{N-i} ~g_<(x,t;y,0) \cr
&~~+ (N-1)~\binom{N-2}{i-1} ~q(x,t)^{i-1}~p(x,t)~[1-q(x,t)]^{N-i-1} ~g_>(x,t;y,0)\Bigg]  \cr
&~~~~~~\times \Bigg[N\binom{N-1}{j-1}~ \big[q(y,0)\big]^{j-1} ~p (y,0)~\big[1-q(y,0)\big]^{N-j}\Bigg],
\end{align}
\begin{align}
 {\rm with}~~    p(x,t)=\int^{\infty}_{-\infty} d\bar x ~g(x,t|\bar x,0)~ \phi (\bar x),~~q(x,t) = \int^{x}_{-\infty}dz~ p(z,t).
\end{align}
$g_<(x,t;y,0)=\int_{-\infty}^{x} dx'~g(x',t|y,0)$ in the first term on the right-hand side of Eq.~\eqref{eq:rxtry0_Sij} corresponds to the case where the $j$th particle, initially at $y$ arrives to the left of the position $x$ at time $t$, while $g_>(x,t;y,0)=\int_{x}^{\infty} dx'~g(x',t|y,0)$ in the second term accounts for the case where it arrives to the right of the position $x$ at time $t$.
Combining \eqref{eq:rxtry0_Rij} and \eqref{eq:rxtry0_Sij} and after some algebraic simplifications, we  get the expression in Eq.~\eqref{eq:conn_corr_jpdf_twotime-full}.

\section{Derivation of Eq.~\eqref{eq:rxryt_expr_full}}\label{sec:appB}
To compute $\langle \hat{\varrho}(X,t) \hat{\varrho}(Y,t)\rangle$, we again transform the hard rod configuration onto the corresponding point particle representation following Eq.~\eqref{eq:HR to HP mapping infinite line}. Within this framework, the joint probability
\begin{align}
    \mathscr{P}_{ij}(x,t;y,t)=\langle \delta(x-x_i(t)) ~ \delta(y-x_j(t)) \rangle
\end{align}
represents the probability that the $i^{\rm th}$ point particle is at $x$ and the $j^{\rm th}$ point particle is at $y$ at same time $t'=t$. The evaluation of $\mathscr{P}_{ij}(x,t;y,t)$ proceeds by considering two possible spatial orderings at time $t,$ the  $j^{\rm th}$ particle located at position $y$ may lie either (i) to the right or (ii) to the left of the $i^{\rm th}$ particle located at position $x$. Focusing on case (i), when the the $j^{\rm th}$ particle lies to the right of the $i^{\rm th}$ particle (i.e., for $x<y$) $\mathscr{P}_{ij}(x,t;y,t)$ can be written as
\begin{subequations}\label{eq:Pxtyt}
    \begin{align}\label{eq:Pxtyt1}
    \mathscr{P}_{ij}(x,t;y,t)=N(N-1)~ \msc R_{ij}(x,y,t)
\end{align}
with
\begin{align}\label{eq:Pxtyt2}
\begin{split}
  \msc R_{ij}(x,y,t)=&\binom{N-2}{i-1} \binom{N-i-1}{j-i-1} ~q(x,t)^{i-1}~ p(x,t)\\
  &~~~~~\times ~\bar q(y,x,t)^{j-i-1}~p(y,t)[1-q(y,t)]^{N-j},  
  \end{split}
\end{align}
\end{subequations}
where $\bar q(y,x,t)=\int_{x}^y dz~p(z,t)$ and the probabilities $p(x,t)$ and $q(x,t)$ are given in Eqs.~\eqref{eq:prob_p}-\eqref{eq:prob_q}. 
In Eq.~\eqref{eq:Pxtyt}, the prefactor $N(N-1)$ accounts for the number of distinct ways of selecting two particles from the initial configuration: one particle can be placed at position $x$ at time $t$ in $N$ possible ways, and a second particle can be placed at position $y$ in $(N-1)$ ways. The factor $\msc R_{ij}(x,y,t)$ ensures the constraint that at time $t$, the particle at $x$ to be the $i^{\rm th}$ and the particle at $y$ to be the $j^{\rm th}$ particle, respectively. An analogous expression can be obtained for the case (ii) where the $j^{\rm th}$ particle lies to the left of the $i^{\rm th}$ particle (i.e., for $x>y$) as $\mathscr{P}_{ij}(x,t;y,t)=N(N-1) \msc R_{ij}(y,x,t)$.
Thus the joint probability $\mathscr{P}_{ij}(x,t;y,t)$ can be written as
\begin{align}\label{eq:Pxyt_full}
\mathscr{P}_{ij}(x,t;y,t)=
\begin{cases}
    N(N-1) \msc R_{ij}(x,y,t)~~~~~~~\text{for~}x<y\\
    N(N-1) \msc R_{ij}(y,x,t)~~~~~~~\text{for~}x>y,
\end{cases}
\end{align}
where $\msc R_{ij}(x,y,t)$ is given in Eq.~\eqref{eq:Pxtyt2}. We substitute the expression of $\mathscr{P}_{ij}(x,t;y,t)$ from Eq.~\eqref{eq:Pxyt_full} into Eq.~\eqref{eq:rXtrYt'_def_pp_full}. Considering the fact that $\langle \hat{\varrho}(X,t) \hat{\varrho}(Y,t)\rangle$ is zero for $|X-Y|<a,$  one can calculate $\langle \hat{\varrho}(X,t) \hat{\varrho}(Y,t)\rangle$ for $|X-Y| \ge a$ as given in Eq.~\eqref{eq:rxryt_expr_full}.

\section{Derivation of Eq.~\eqref{eq:largeNrxryt_1}}
\label{sec:appB1}
In Fourier space, by substituting the expression of $\langle \hat{\varrho}(X,t) \hat{\varrho}(Y,t)\rangle$ from Eq.~\eqref{eq:rxryt_expr_full} into Eq.~\eqref{eq:rxryt_FFT}, the joint distribution of the mass density can be written as 
\begin{align}
    \tilde {\mt P}(k_1,k_2,t)&= \int_{-\infty}^{\infty}dz_1 \int_{-\infty}^{\infty}dz_2~e^{\iota k_1z_1}e^{\iota k_2z_2}~\bar \rho(z_1, t)~\bar \rho(z_2-a,t)\cr
     &~~\times ~\left[1+\left(e^{\iota (k_1+k_2)a}-e^{\iota k_2a}\right)\frac{\bar F(z_1,t)}{N}+\left(e^{ik_2a}-1\right)\frac{\bar F(z_2-a,t)}{N}\right]^{N-2},\label{eq:largeNrxry1_app}
\end{align}
where we consider $\bar \rho(z,t)=Np(z,t),\bar F(z,t)=Nq(z,t).$ To get an approximate expression of the distribution function for large but finite $N,$ we rewrite the term $[\dots]^{(N-2)}$ term in Eq. (\ref{eq:largeNrxry1_app}) as
\begin{align}
    &\left[1+\left(e^{\iota (k_1+k_2)a}-e^{\iota k_2a}\right)\frac{\bar F(z_1,t)}{N}+\left(e^{\iota k_2a}-1\right)\frac{\bar F(z_2-a,t)}{N}\right]^{N-2}\cr
    &=\exp\Bigg[(N-2) \log \Big[1+\left(e^{\iota (k_1+k_2)a}-e^{\iota k_2a}\right)\frac{\bar F(z_1,t)}{N}+\left(e^{\iota k_2a}-1\right)\frac{\bar F(z_2-a,t)}{N}\Big]\Bigg]\nonumber\\ \label{eq:largeNrxry2_app}
\end{align}
We next expand the logarithmic term on the right-hand side of Eq.~\eqref{eq:largeNrxry2_app} for large $N$ and small $a$, keeping $\frac{Na}{\sigma}$ finite. Approximating $e^{\iota ka}=1+\iota ka-\frac{k^2a^2}{2}+{\mathcal O}(k^3a^3),$ and retaining terms up to quadratic order in $k_1,k_2$ in the limit $N\to\infty$ we get
\begin{align}
    \tilde {\mt P}(k_1,k_2,t)&=\int_{-\infty}^{\infty}dz_1 \int_{-\infty}^{\infty}dz_2~e^{\iota k_1z_1}e^{\iota k_2z_2}~\bar \rho(z_1, t)~\bar \rho(z_2-a,t)\cr
 &~~~\times~\exp\Bigg[\iota k_1a\bar F(z_1,t)-\frac{k_1^2a^2}{2N}\Sigma_{1a}^2(z_1,t)+\iota k_2a\bar F(z_2-a,t) - \frac{k_2^2a^2}{2N}\Sigma_{2a}^2(z_2,t)\cr
 &~~~~~~~~~~~~~~~~~~~-\frac{k_1k_2a^2}{N}\Sigma_{a}^2(z_1,z_2,t)\Bigg],
\end{align}
where $\Sigma_{1a}^2(z_1,t)=\bar F(z_1,t)\Big[N-\bar F(z_1,t)\Big],$ $\Sigma_{2a}^2(z_2,t)= \bar F(z_2-a,t)\Big[N-\bar F(z_2-a,t)\Big]$ and $\Sigma_{a}^2(z_1,z_2,t)=\bar F(z_1,t)\Big[N-\bar F(z_2-a,t)\Big].$ Performing the inverse Fourier transform of $\tilde{\varrho}(k_1,k_2,t)$, we obtain
\begin{gather}\label{eq:largeN_rhox1x2_app}
\begin{split}
     \langle \hat{\varrho}(X,t) \hat{\varrho}(Y,t)\rangle =& \int_{-\infty}^{\infty}dz_1 \int_{-\infty}^{\infty}dz_2 \int_{-\infty}^{\infty} \frac{dk_1}{2\pi}  \int_{-\infty}^{\infty} \frac{dk_2}{2\pi}~ \bar \rho(z_1, t)~\bar \rho(z_2-a,t) \\& 
     \times~\exp\Bigg[-\iota k_1(X-z_1- a\bar F(z_1,t))-\iota k_2(Y-z_2-a\bar F(z_2-a,t)) \\&~~- \frac{k_1^2a^2}{2N}\bar F(z_1,t)\big[N-\bar F(z_1,t)\big]  - \frac{k_2^2a^2}{2N}\bar F(z_2-a,t)\big[N-\bar F(z_2-a,t)\big] \\&
 ~~-\frac{k_1k_2a^2}{N}\bar F(z_1,t)\big[N-\bar F(z_2-a,t)\big]\Bigg]
\end{split}
\end{gather}
To evaluate the integrals over $k_1$ and $k_2$ in Eq.~\eqref{eq:largeN_rhox1x2_app}, we employ the standard multidimensional Gaussian identity
\begin{gather}\label{eq:Gaussian_linear}
    \int_{-\infty}^{\infty} (d \mathbf{x}) \exp\big[-\frac{1}{2} (\mathbf{x},\mathbf{Ax})+\mathbf{(b,x)}+c\big] = \frac{(2\pi)^\frac{n}{2}}{\sqrt{\det(\mathbf{A})}} \exp\big[\frac{1}{2} \mathbf{(b,A^{-1}b)}-c \big]
\end{gather}
where $\mathbf{x}$ and $\mathbf{b}$ are multidimensional vectors, $\mathbf{A}$ is an $n\times n$ symmetric matrix, $(d\mathbf{x})=dx_1 dx_2...dx_n$ and $c$ is a scalar. In the present case, corresponding to Eq.~\eqref{eq:largeN_rhox1x2_app}, we identify the transpose of the vector $\mathbf{k},$ $\mathbf{k}^T=\big(k_1,k_2\big)$, $c=0$ and
\begin{gather}
\begin{split}
   &\mathbf{b}^T = \Big( -\iota(X-z_1- a\bar F(z_1,t)),\ -\iota(Y-z_2-a\bar F(z_2-a,t)) \Big) \cr
   &\mathbf{A} = \frac{a^2}{N} \begin{pmatrix} \bar F(z_1, t) \big( N - \bar F(z_1, t) \big) & \bar F(z_1, t) \big( N - \bar F(z_2 - a, t) \big) \\
\bar F(z_1, t) \big( N - \bar F(z_2 - a, t) \big) & \bar F(z_2 - a, t) \big( N - \bar F(z_2 - a, t) \big) \end{pmatrix}.
\end{split}
\end{gather}
Thus one can compute $\langle \hat{\varrho}(X,t) \hat{\varrho}(Y,t)\rangle$ as
\begin{gather}\label{eq:largeNrxryt_2_app}
    \langle \hat{\varrho}(X,t) \hat{\varrho}(Y,t)\rangle = \frac{1}{(2\pi)^2}\int_{-\infty}^\infty dz_1 \int_{-\infty}^\infty dz_2 ~ \bar \rho(z_1, t) \bar \rho(z_2-a, t) \frac{1}{ \sqrt{\det(\mathbf A)}} \exp\left(-\frac{1}{2} \mathbf{Y}^T \mathbf A^{-1} \mathbf{Y}\right)
\end{gather}
with
\begin{equation}
    \mathbf{Y} = \begin{pmatrix} X - \mu_1 \\ Y - \mu_2\end{pmatrix}, \hspace{0.5cm} \mu_1= z_1 + a \bar F(z_1, t), \hspace{0.3cm} \mu_2=z_2 + a \bar F(z_2-a, t).
\end{equation}

\section{BMFT of hydrodynamic scale correlation of mass density}
\label{sec:hrdro-corr}
In this section we compute the hydrodynamic scale correlation of the coarse-grained mass density $\varrho(\mt X, \mtt)$ defined  in Eq.~\eqref{def:hd_corr}. We start with Eq.~\eqref{def:hd-corr-pathInt}, in which we first rewrite the probability functional $\mc P_{\rm r}[\mf f(\mt X,v,\mtt)]$ in the following way.
Using the integral representations of the delta functional in Eq.~\eqref{mcP[mff(t)]} by introducing an auxiliary field $\mf  h(\mtX,v,\mtt)$ and the delta function in Eq.~\eqref{mcP[mff(0)]} by introducing a chemical potential $\mu$, we can rewrite the expression of correlation $\mc C$ in Eq.~\eqref{def:hd-corr-pathInt} as \cite{AnupamMFT}
\begin{subequations}
\label{eq:path_int}
\begin{align}
\begin{split}
&\mc C(\ell\mt X_a, \ell\mtt_a;\ell\mt X_b, \ell\mtt_b)  \\
&~=- \frac{1}{\ell}\left [ \frac{\partial}{\partial (\lambda/\ell)} 
\left \{ \frac{\int d\mu~e^{\mu N}\int \mc D[\mf f(\mtX, v,\mtt), \mf  h(\mtX, v,\mtt)]\varrho(\mtX_b,\mtt_b)e^{-\ell \mc S^{\rm r}_{\frac{\lambda}{\ell}}[\mf f(\mtX, v,\mtt), \mf  h(\mtX, v,\mtt)] }}{\int d\mu~e^{\mu N}\int \mc D[\mf f(\mtX, v,\mtt), \mf  h(\mtX, v,\mtt)]e^{-\ell \mc S_{\frac{\lambda}{\ell}}[\mf f(\mtX, v,\mtt), \mf  h(\mtX, v,\mtt)] }}\right\}
\right]_{\lambda=0}
\end{split}
\label{eq:path_int1}
\end{align}
where,
\begin{align}
\begin{split}
\mc S_\lambda^{\rm r}[\mf f, \mf h]=&~\mc F^{\rm r}_0[\mf f(0)]- \mc F^{\rm r}_{0}[\mf f_{\rm eq}] + \mu \int d\mt Y \int du ~\mf f(\mt Y, u,0) \\
&+\lambda \int_0^{\mt T} d\mtt \int d\mt Y \int du~ \delta(\mtt -\mtt_a)\delta(\mt Y-\mt X_a) \mf f(\mt Y, u, \mtt)  \\
&+ \int_0^{\mt T} d\mtt \int d\mt Y \int du ~\mf  h(\mt Y, u, \mtt) 
\{\partial_{\mtt} \mf f(\mt Y, u, \mtt) +\partial_{\mt Y}v_{\rm eff}[\mf f]\mf f(\mt Y,v,\mtt) \},
\end{split}
\label{eq:action}
\end{align}
\end{subequations}
where recall $\mf f_{\rm eq}(\mt Y, u, 0)$ corresponds to the equilibrium profile that minimizes the free energy $\mc F^{\rm r}[\mf f(0)]$. Note that the expression of the hydro-scale correlation in Eq.~\eqref{eq:path_int1} is already in the scaling form in Eq.~\eqref{eq:corr_scaling}. To find the scaling function $\mt C(\mt X_a,\mtt_a;\mt X_b,\mtt_b)$, our next task is to perform  the path integrals, both in the numerator and the
denominator of Eq.~\eqref{eq:path_int1},  using saddle point method and find the saddle point density as mentioned in Eq.~\eqref{eq:scaled-corr-rho_sd}. 

In many cases it seems convenient to convert  the saddle point problem in hard rod picture to an equivalent saddle point problem in point particle picture where one minimizes an appropriate action written in terms of the point particle phase space distribution $\mc  f(\mt x, u,\mtt)$ and finds the saddle point mass density $\rho^{\rm sd}_\lambda(\mt x,\mtt)= \int du~\mc  f^{\rm sd}_\lambda(\mt x,u,\mtt)$. The saddle point density $\varrho_\lambda^{\rm sd}(\mt X,\mtt)$ of hard rods is related to that of point particles   as 
\begin{subequations}
\label{eq:macro_rholambda_transform0}
\begin{align}\label{eq:macro_rholambda_transform}
  \varrho^{\rm sd}_{\lambda}(\mt X_{\mc f^{\rm sd}_\lambda(\mtt)}(\mt x),\mt t)=\frac{\rho^{\rm sd}_{\lambda}(\mt x,\mt t)}{(1+a\rho^{\rm sd}_{\lambda}(\mt x,\mt t))},
\end{align}
where
\begin{align}\label{eq:macro_X_transform}
\mt X_{\mc f(\mtt)}(\mt x)= \mt x + a\int d\mt z \int du~\Theta(\mt x-\mt z) \mc  f(\mt z, u, \mtt),
\end{align}
\end{subequations} 
defines the transformation from point particle coordinates to hard rod coordinates. The inverse transformation 
\begin{align}\label{eq:macro_x_transform}
\mt x_{\mf f(\mtt)}(\mt X)= \mt X - a\int d\mt Z \int du~\Theta(\mt X-\mt Z) \mf f(\mt Z, u, \mtt),
\end{align} 
defines the point particle  coordinates in terms of the hard rod phase space density $\mf  f(\mt Z, u,\mtt)$ at (scaled) time $\mtt$.

\subsection{Saddle point density $\mc{f}^{\rm sd}_\lambda(\mt x, u,\mtt)$}
In the point particle picture,  one needs to minimise the following action [see \cite{AnupamMFT}]
\begin{align}
\begin{split}
\mc S_\lambda^{\rm p}[\mc f,\mc h]=&~ \mc F^{\rm p}[\mc  f(0)]- \mc F^{\rm p}[\mc  f_{\rm eq}] + \mu \int d\mt y \int du ~\mc  f(\mt y, u,0) \\
&+ \lambda \int_0^{\mt T} d\mtt \int d\mt y \int du~ \delta(\mtt -\mtt_a)~\delta(\mt X_{\mc  f(\mtt)}(\mt y)-\mt X_a) \mc  f(\mt y, u, \mtt)  \\
&+ \int_0^{\mt T} d\mtt \int d\mt y \int du ~\mc   h(\mt y, u, \mtt) 
\{\partial_{\mtt} \mc  f(\mt y, u, \mtt) +u\partial_{\mt y}\mc  f(\mt y,v,\mtt) \},
\end{split}
\label{S^p[f(t)]}
\end{align}
where 
\begin{align}
\mc F^{\rm p}[\mc  f(0)] = \int d\mt z \int du~\mc  f(\mt z,u,0)\left[ \mc g(\mt z,u)+ \ln \mc  f(\mt z,u,0)\right], \label{F^p[f(0)]}
\end{align}
 with $\mc g(\mt z,u)= -\ln  \phi_0(\mt z) - \ln h(u) $ [see discussion above Eq.~\eqref{def:mfb_f}].
Minimising the action in Eq.~\eqref{S^p[f(t)]}, we get the following saddle point equations \cite{AnupamMFT}
\begin{subequations}
\label{eq:saddle}
    \begin{align}
        &\partial_{\mt t} \mc  f^{\rm sd}_\lambda(\mt x,v,\mt t)+v\partial_{\mt x} \mc  f^{\rm sd}_\lambda(\mt x,v,\mt t)=0, \label{eq:sad-1}\\
        &\partial_{\mt t} \mc  h^{\rm sd}_\lambda(\mt x,v,\mt t)+v\partial_{\mt x} \mc  h^{\rm sd}_\lambda(\mt x,v,\mt t)=\lambda \delta(\mtt-\mtt_a)\mc B(\mt x, v), \label{eq:sad-2}\\
        &\mc  h^{\rm sd}_\lambda(\mt x,v,\mt T)=0, \label{eq:sad-3}\\
        &\mc  h^{\rm sd}_\lambda(\mt x,v,0)=\ln \frac{\mc  f^{\rm sd}_{\lambda}(\mt x,v,0)}{\mc f_{\rm eq}(\mt x, v)}+\mu, \label{eq:sad-4}
    \end{align}
    defined over the time domain $0\leq \mtt \leq \mt T$ with $\mt T> \max(\mtt_a,\mtt_b)$, where 
    \begin{align}\label{eq:Bxv_app}
    \mc B(\mt x, v) =& \delta(\mt X_{\mc  f^{\rm sd}_\lambda(\mtt)}(\mt y)-\mt X_a)+a\int d\mt y \int du~ 
    \{\partial_{\mt X_{\mc f^{\rm sd}_\lambda(\mtt)}(\mt y)} \delta(\mt X_{\mc  f^{\rm sd}_\lambda(\mtt)}(\mt y)-\mt X_a)\}~\notag \\ 
    &~~~~~~~~~~~~~~~~~~~\times~\mc f^{\rm sd}_\lambda(\mt y, u,\mt t)~\Theta(\mt y-\mt x), \notag \\
    =& \partial_{\mt X_{\mc f^{\rm sd}_\lambda(\mtt)}(\mt x)} \left[(1-a \mf \varrho^{\rm sd}_\lambda(\mt X_{\mc f^{\rm sd}_\lambda(\mtt)}(\mt x),\mtt_a))\Theta(\mt X_a-\mt X_{\mc f^{\rm sd}_\lambda(\mtt)}(\mt x)) \right],
    \end{align}
    with
    \begin{align}
        \mc f_{\rm eq}(\mt x,v) = Ne^{-\mc g(\mt x,v)}=N \phi_0(\mt x)h(v).
    \end{align}
The saddle point equations in ~\eqref{eq:saddle} are also known as the BMFT equations in the literature \cite{doyon2023ballistic, AnupamMFT}. To determine $\mc{f}^{\rm sd}_\lambda(\mt x, u,\mtt)$, the last step is to solve the above equations with appropriate boundary conditions. The constant $\mu$ is fixed from  the normalization $N=\int d\mt z \int du~\mc f^{\rm sd}_\lambda(\mt z, u,0)$. Note if the initial phase space density satisfies the normalization condition, the density at later time will also satisfy the normalization by virtue  of the continuity equation \eqref{eq:sad-1}. 
In the next section, we discuss the solution of the BMFT equations in detail and using it, we compute the two-point correlation $\mt C(\mt X,\mtt;\mt Y,\mtt')$. For notational convenience, we omit the subscript $(...)^{\rm sd}$ in the subsequent calculations.
\end{subequations}

\subsection{Derivation of  $\mt C(\mt X,\mt t; \mt X_a,\mtt_a)$}
\label{sec:appC}
In this section, to compute the correlation $\mt C(\mt X,\mt t; \mt X_b,\mtt_b)$ we solve the saddle point equations in Eq.~\eqref{eq:saddle}. We first integrate both sides of Eq.~\eqref{eq:sad-2} over time form $(\mt t_a-\epsilon)$ to $(\mt t_a+\epsilon)$ with $\epsilon >0$ and get the value of $\mc  h_{\lambda}(\mt x,v,\mt t_a^{-})=\mc  h_{\lambda}(\mt x,v,\mt t_a-\epsilon)|_{\epsilon \to 0}$ as
\begin{align}
    \mc  h_{\lambda}(\mt x,v,\mt t_a^{-})=-\lambda~ \mathcal{B}(\mt x,v).
\end{align}
Note that here $\mc h_{\lambda}(\mt x,v,\mt t > \mt t_a)=0$, which can be obtained by evolving Eq.~\eqref{eq:sad-2} backward in time from $\mt T$ to $\mt t_b$ and considering $\mc  h_{\lambda}(\mt x,v,\mt T)=0$ following Eq.~\eqref{eq:sad-3}. Then we evolve Eq.~\eqref{eq:sad-2} from $\mt t=\mt t_a^{-}$ to $\mt t=0$ backward and get 
\begin{align}
    \mc  h_{\lambda}(\mt x,v,0)=\mc  h_{\lambda}(\mt x+v\mt t_a,v,\mt t_a)=-\lambda~ \mathcal{B}(\mt x+v\mt t_a,v).
\end{align}
Substituting $\mc  h_{\lambda}(\mt x,v,0)$ in the lhs of Eq.\eqref{eq:sad-4} and solving one finds 
\begin{align}
    \mc f_\lambda(\mt x,v,0) =  \frac{\mc f_{\rm eq}(\mt x,v)e^{-\lambda \mc B(\mt x+v \mt t_a,v)}}{\int d\mt z \int du~ \mc g(\mt z,u)e^{-\lambda \mc B(\mt z+u \mt t_a,u)}},
\end{align}
where the value of $\mu$ has been  fixed by normalization $\int d\mt x \int dv~\mc f_\lambda(\mt x,v,0)=N$ and  $\mc g(\mt z,u) =- \ln(\phi_0(\mt z)h(u))$. Evolving $\mc  f_{\lambda}(\mt x,v,0)$ forward in time to $\mt t$ we obtain 
\begin{align}
    \mc  f_{\lambda}(\mt x,v,t)=\mc  f_{\lambda}(\mt x-v\mt t,v,0) =  \frac{\mc f_{\rm eq}(\mt x-v \mtt,v)e^{-\lambda \mc B(\mt x+v (\mt t_a-\mtt),v)}}{\int d\mt z \int du~\mc g(\mt z,u)e^{-\lambda \mc B(\mt z+u \mt t_a,u)}}.
\end{align}
Inserting this solution in Eq.~\eqref{eq:macro_rholambda_transform0} with $\rho_\lambda(\mt x) = \int dv~\mc f_\lambda(\mt x, v,\mt t)$, one finds 
\begin{subequations}
\begin{align}
\varrho_{\lambda}(\mt X,\mt t)=\frac{\rho_{\lambda}(\mt x_{\mf f_{\lambda}(\mtt)}(\mt X),v,\mt t)}{(1+a\rho_{\lambda}(\mt x_{\mf f_{\lambda}(\mtt)}(\mt X),\mt t))},
\end{align}
where, recall
\begin{align}
    \mt x_{\mf f_{\lambda}(\mtt)}(\mt X)= \mt X - a\int d\mt Z \int du~\Theta(\mt X-\mt Z) \mf  f_{\lambda}(\mt Z, u, \mtt).
\end{align}
\end{subequations}
Finally, evaluating the derivative  $\partial_\lambda \varrho_\lambda(\mt X_b,\mtt_b)$ at $\lambda =0$, we get the scaled correlation function  $\mt C(\mt X_a,\mtt_a;\mt X_b,\mtt_b)$ in general for $\mtt_a\ne \mtt_b$.
For $\mtt_a=\mtt_b$, the expression simplifies and we get the equal time correlation as announced in Eq.~\eqref{eq:corr_expr_bmft_full} of the main text. For the un-equal time correlation $\mt C(\mt X_a, t_a;\mt X_b,0)$, we put $t_b=0$ and get the expression given in Eq.~\eqref{eq:twotime_bmft_full}. Note that both correlations in Eq.~\eqref{eq:corr_expr_bmft_full} and Eq.~\eqref{eq:twotime_bmft_full} 
are expressed in terms of the solution $\bar \varrho(\mt X, \mtt) = \varrho_{\lambda \to 0}(\mt X, \mtt)$ of the Euler GHD. This solution is obtained in Eq.~\eqref{sol:bar_varrho_Eu}.  
\section*{References}

\bibliographystyle{unsrt}
\bibliography{references}
\end{document}